\documentclass[mnsc,nonblindrev]{informs3_nojournal}

\OneAndAHalfSpacedXI

\usepackage{natbib}
 \bibpunct[, ]{(}{)}{,}{a}{}{,}%
 \def\bibfont{\small}%
 \def\bibsep{\smallskipamount}%
 \def\bibhang{24pt}%
 \usepackage[section]{placeins}

\TheoremsNumberedThrough     
\ECRepeatTheorems

\EquationsNumberedThrough    

\MANUSCRIPTNO{}

\usepackage[utf8]{inputenc} 
\usepackage[T1]{fontenc}    
\usepackage{hyperref}       
\usepackage{url}            
\usepackage{booktabs}       
\usepackage{amsfonts}       
\usepackage{nicefrac}       
\usepackage{microtype}      
\usepackage[flushleft]{threeparttable}
\usepackage{comment}

\usepackage{multibib}
\newcites{RR}{Response References}
\usepackage{subcaption}
\usepackage{booktabs}
\usepackage{graphicx}
\usepackage{mathtools}
\usepackage{hyperref}
\usepackage{xcolor}
\usepackage[flushleft]{threeparttable}
\usepackage[linesnumbered,algoruled,boxed,lined]{algorithm2e}

\usepackage{amsmath}
\usepackage{amssymb}
\usepackage{float}

\usepackage{titlesec}
\usepackage[T1]{fontenc}
\usepackage{lmodern}

\newcommand{\edit}[1]{\textcolor{black}{#1}}
\newcommand{\blockedit}{\color{black}}

\titleformat{\subsubsection}
{\normalfont\normalsize\itshape}{\thesubsubsection}{1em}{}
\titlespacing*{\subsubsection}{0pt}{1.75ex plus 1ex minus .2ex}{0ex plus .2ex}

\begin{document}


\RUNTITLE{Market Segmentation Trees}
\RUNAUTHOR{Aouad, Elmachtoub, Ferreira, McNellis}
\TITLE{Market Segmentation Trees}

\ARTICLEAUTHORS{%
\AUTHOR{Ali Aouad}
\AFF{London Business School, London, UK, \EMAIL{aaouad@london.edu}} 
\AUTHOR{Adam N. Elmachtoub}
\AFF{Department of Industrial Engineering and Operations Research and Data Science Institute, Columbia University, New York, NY, \EMAIL{adam@ieor.columbia.edu}}
\AUTHOR{Kris J. Ferreira}
\AFF{Harvard Business School, Harvard University, Boston, MA, \EMAIL{kferreira@hbs.edu}}
\AUTHOR{Ryan McNellis}
\AFF{Amazon\footnote{Work completed prior to joining Amazon.}, New York, NY, \EMAIL{rmcnell@amazon.com}}
} 

\ABSTRACT{%
\textbf{Problem Definition:} We seek to provide an interpretable framework for segmenting users in a population for personalized decision-making. 
 
\textbf{Methodology / Results:} We propose a general methodology, \textit{Market Segmentation Trees} (MSTs), for learning market segmentations explicitly driven by identifying differences in user response patterns. To demonstrate the versatility of our methodology, we design two new, specialized MST algorithms: $(i)$ Choice Model Trees (CMTs), which can be used to predict a user's choice amongst multiple options and $(ii)$ Isotonic Regression Trees (IRTs), which can be used to solve the bid landscape forecasting problem.
We provide a theoretical analysis of the asymptotic running times of our algorithmic methods, which validates their computational tractability on large datasets. We also provide a customizable, open-source code base for training MSTs in Python which employs several strategies for scalability, including parallel processing and warm starts. Finally, we assess the practical performance of MSTs on several synthetic and real world datasets, showing that our method reliably finds market segmentations which accurately model response behavior. 

\textbf{Managerial Implications:} 
The standard approach to conduct market segmentation for personalized decision-making is to first perform market segmentation by clustering users according to similarities in their contextual features, and then fit a ``response model'' to each segment in order to model how users respond to decisions. 
\edit{However, this approach may not be ideal if the contextual features prominent in distinguishing clusters are not key drivers of response behavior. Our approach addresses this issue by integrating market segmentation and response modeling, which consistently leads to improvements in response prediction accuracy, thereby aiding personalization. We find that such an integrated approach can be  computationally tractable and effective even on large-scale datasets. Moreover, MSTs are interpretable since the market segments can easily be described by a decision tree and often require only a fraction of the number of market segments generated by traditional approaches.}
}%



\maketitle

\section{Introduction}

Recent growth of e-commerce and media consumption have resulted in an expansion of opportunities for firms to engage in personalized decision-making. Online retailers often offer product recommendations on their homepage, which are personalized using the visiting user's purchase history and demographic information. Streaming services such as YouTube and Spotify personalize ads based on the media content being consumed and other aspects of the user's activity history. Online search engines such as Google personalize the ranking of search results based on user  history and location. 

Personalized decision-making often lies at the intersection of two fundamental technical challenges:
\textit{market segmentation} (clustering users into segments based on user characteristics) and \textit{response modeling} (the probabilistic modeling of a user's response to a personalized decision). For example, if an online platform wishes to personalize the ads displayed to its users in order to maximize the click-through rate, it could (1) partition users into disjoint segments, and (2) model the click behavior of users in each segment. The standard approach is to perform the tasks of market segmentation and response modeling separately, using a clustering algorithm (e.g., $K$-means) for market segmentation and then fitting a response model (e.g., logistic regression) within each cluster \citep{yang2016buyer}. \edit{However, such a market segmentation is driven only by user feature dissimilarity rather than differences in user response behavior, which can be particularly problematic  in cases where the key user features distinguishing market segments are not key drivers of response behavior.}


We propose a general methodology, Market Segmentation Trees (MSTs), that builds interpretable decision trees for \textit{joint} market segmentation and response modeling, which can be used for a variety of personalized decision-making applications. Decision tree splits are applied by the MST to segment the market according to available \textit{contextual} attributes for personalization (e.g., features encoding the user). The leaves of the tree correspond to market segments. A response model is fit in each segment to  model the users' response (e.g., clicks) probabilistically as a function of the decision variables (e.g., ads to be displayed or product prices). \edit{We propose a training procedure for MSTs in which decision tree splits are decided by optimizing the predictive accuracy of the resulting collection of response models.} Thus, our training procedure yields a market segmentation driven by accurately capturing differences in user response behavior. 

We emphasize that a primary motivation for the use of decision trees for tackling this problem is due to their interpretability, in addition to their strong predictive performance. Increasingly, companies are being held more accountable for their data-driven decisions by both consumers and regulators \citep{goodman2017european}. Decision trees provide a simple way to visualize the decision-making stream and have been used in a variety of settings \citep{kallus2017recursive, elmachtoub2017practical, ciocan2020interpretable, bertsimas2019optimal, aghaei2019learning}. 
\edit{In our setting, interpretability stems from the fact that every user is mapped to a single response model simply by observing where the user's contextual attributes fall on the tree, and each potential response model is a function solely of the decision variables.}

We provide an open-source implementation of our training procedure in Python \footnote{\url{https://github.com/rtm2130/MST}}. The code base is modular and easily customizable to fit different personalized decision-making applications. Several features have been included for improved scalability, including the option of using parallel processing and warm starts for training the MST models. We provide a theoretical analysis of the code's asymptotic computational complexity supporting its tractability in large data settings. Specifically, we show that under mild conditions, the implementation's computational complexity is \textit{linear} in the depth of the learned MST; moreover, the impact of tree depth on computational complexity can be greatly diminished if a sufficient number of cores are available for parallel processing.


To demonstrate the versatility of our methodology, we design two specialized MST algorithms. First, we propose \textit{Choice Model Trees} (CMTs), which can be used to predict a user's choice among multiple options. Our model uses decision tree splits to segment users on the basis of their features (e.g., prior purchase history), and within each segment/leaf a Multinomial Logit (MNL) choice model is fit as the response model to predict the probability that users in that segment choose each option. We examine the performance of CMTs on a variety of synthetic datasets, observing that CMTs reliably find market segmentations which accurately predict choice probabilities, whereas other natural benchmarks do not. Furthermore, we show that CMTs are more easily able to overcome model misspecification and are quite robust to overfitting. \edit{Next, we apply the CMT to the Swissmetro dataset \citep{bierlaire2001acceptance}, which captures how people choose various transportation options based on user and option features. We find that CMTs can outperform \edit{ a variety of benchmarks}, while providing an interpretable segmentation of users. 
}

We also propose a second algorithm derived from our MST framework, \textit{Isotonic Regression Trees} (IRTs), which can be used to solve the bid landscape forecasting problem under first-price auction dynamics; most major ad exchanges are anticipated to switch to first-price auctions \citep{sluis2019google}. A ``bid landscape'' refers to the probability distribution of the highest (outside) bid that an ad spot will receive when being auctioned at an advertising exchange. The bid landscape forecasting problem is important to Demand Side Platforms (DSPs) -- ad campaign management platforms --  in estimating the minimum bid necessary to win different types of ad spots. 
Our model uses a decision tree to segment auctions according to features about the visiting user (e.g., user's location) and the ad spot being auctioned (e.g., width/height in pixels). An isotonic regression model is used as the response model to forecast the bid landscapes of the auctions within each segment. IRTs are fully non-parametric, operating without assumptions about the distribution of the bid landscapes or of their relationship with the auction features. We apply our IRT to an ad spot transaction dataset collected by a large DSP provider, and we demonstrate that our model  achieves a 5-29\% improvement in bid landscape forecasting accuracy over the DSP's current approach across multiple ad exchanges, while requiring significantly fewer market segments than traditional approaches. 

\section{Literature Review}

In this work, we propose a general framework (MSTs) for building decision trees for integrating market segmentation and personalized decision-making. An introduction to decision trees may be found in \citet{friedman2001elements}, \edit{as well as more recent surveys on their interpretability \cite{rudin2019stop} and optimization algorithms \citep{carrizosa2021mathematical}.} MSTs take the structural form of model trees, which refer to a generalization of decision trees that allow for non-constant leaf prediction models. The most common model tree algorithms explored in the literature are linear model trees \citep{quinlan1992learning} and logistic model trees \citep{chan2004lotus,landwehr2005logistic}, which propose using linear and logistic regression leaf models with decision trees. \citet{zeileis2008model} develop a general framework, model-based recursive partitioning (MOB), for training model trees with parametric leaf models such as linear and logistic regression. Unlike our training methodology, none of the above methods select decision tree splits which \textit{directly} minimize the predictive error of the resulting collection of leaf models, instead employing proxy splitting criteria such as class purity \citep{chan2004lotus,landwehr2005logistic} and parameter instability \citep{zeileis2008model}. We believe this is due to a presumed computational intractability, but we are able to demonstrate that our model trees may be tractably trained in Section \ref{sec:compcomplex}. 

Similar to our CMT algorithm, \citet{mivsic2016data} proposes using model trees with choice model leaves for personalizing assortment decisions. In contrast, MSTs offer a more general framework for building model trees for market segmentation in areas outside of assortment optimization. Moreover, we develop an open-source implementation, which has been empirically validated on  large-scale real-world datasets. \citet{kallus2017recursive} and \citet{bertsimas2019optimal} propose methodology for training decision trees for segmenting customers and personalizing treatments across the resulting segments. Each treatment option is associated with an unknown and customer-variant expected reward, and the authors provide recursive partitioning and integer programming strategies for training the trees to maximize the rewards from the prescribed treatments. The treatment options are assumed to belong to a small set of feasible values and 
thus response models are not needed.
In contrast, MSTs 
support continuous and high-dimensional decision spaces by way of response models. 

\edit{Absent the market segmentation objective, the use of decision trees has recently become quite prevalent in various streams of literature in operations and marketing. \citet{bertsimas2020from}, \citet{cui2018operational},  \citet{ferreira2016analytics}, \citet{glaeser2019optimal}, and \citet{nambiar2019dynamic} use regression tree ensembles (e.g., bagging, boosting, or random forests) to predict demand in retail settings. \citet{lemmens2006bagging} uses classification trees to predict customer churn. \citet{elmachtoub2020decision} trains decision trees for use in the predict-then-optimize framework \citep{elmachtoub2021smart}, which requires modifying the loss function to an optimization problem (rather than a choice model as in this work). \citet{yoganarasimhan2020search} uses boosted regression trees to predict the relevance of a result to a customer's search query. 
\citet{chen2019decision} and \citet{chen2019use} develop new tree-based choice models where each customer type is represented by a decision tree that characterizes the customer's choice given an assortment of available products. 
\citet{mivsic2020optimization} and \citet{biggs2018optimizing} consider how to optimize the input features of tree ensembles in order to maximize the predicted value. 
}

\edit{For many applications, though, market segmentation is an important objective to be considered alongside response modeling.}
The typical approach in industry is to perform the tasks of market segmentation and response modeling separately, first clustering users according to closeness in their contextual attributes and then fitting response models within each cluster \citep{yang2016buyer}. A popular method for doing so is $K$-means clustering -- an unsupervised machine learning algorithm which attempts to find the clustering of users that minimizes the variance of the contextual features within in each cluster. $K$-means clustering is widely utilized for the purposes of market segmentation -- \citet{tuma2011survey} found that $K$-means clustering was the most frequently used market segmentation approach across 210 research articles applying clustering methods for market research (44.25\% of all articles). 
\edit{Unsupervised, tree-based clustering methods have also been proposed for market segmentation, where the resultant tree splits define the clusters, and the loss function used to build the tree reflects the closeness of attributes within each cluster (see, e.g., \citet{blockeel2000topdown}, \citet{fraiman2013interpretable}, and \citet{bertsimas2020interpretable}).}
We argue that the \edit{approach of first clustering and then fitting response models within each cluster}
suffers from a fundamental limitation -- namely, the resulting market segmentation does not take into account the predictive accuracy of the resulting collection of response models but is instead driven only by minimizing within-cluster feature dissimilarity. 

There have been several non-tree-based approaches proposed in the literature for jointly performing market segmentation and response modeling. One of the most popular approaches is the latent-class multinomial logit model (LC-MNL) originally proposed by \citet{kamakura1989probabilistic}. The LC-MNL model creates $K$ different market segments, each one with a separate MNL for modeling response behavior; customers are assigned segment-membership probabilities which may be a function of customer-specific features. MSTs provide a more interpretable market segmentation in that each user is in exactly one segment, rather than probabilistically in each segment. Finally, LC-MNL models are typically fit using Expectation-Maximization (EM) methods which are known to be prohibitively slow on large datasets \citep{jagabathula2020conditional}.





\citet{bernstein2019dynamic} and \cite{kallus2016revealed} propose  \textit{dynamic} market segmentation approaches which adaptively adjust customer segments and their associated response models (Dirichlet Process Mixture and Low Rank Conditionally MNL, respectively)  as more observations are collected.  \citet{yang2016buyer} adapt the $K$-means algorithm to jointly perform market segmentation with response modeling, referring to their approach as \textit{K-Classifiers Segmentation}. These methodologies assume customers are from a predefined set of types, and the algorithms then map the customer types to clusters. \citet{baardman2017leveraging}  propose retroactively fitting a classification machine learning model (e.g., logistic regression) for mapping product features to the cluster assignments outputted by a $K$-Classifiers Segmentation method. In comparison to the aforementioned methods, the MST approach \textit{directly} utilizes available contextual attributes when learning its market segmentation. \citet{jagabathula2018model} propose a method for simultaneous market segmentation and response modeling which (1) fits a response model to the entire population of customers, and (2) segments customers according to how their response behavior differs from the population model (e.g., through a log-likelihood score). This approach does not segment customers on the basis of their demographic features, but rather on their observed historical response behavior. Therefore, their methodology is specialized for personalizing recommendations to \textit{returning} customers, whereas our approach may also be used for personalizing decisions to new customers. 

\section{Methodology}

\subsection{Problem Formulation}
We now provide a general formulation of a personalized decision-making problem, which we break down into three components. First, features $x \in \mathcal{X}\subseteq \mathbb{R}^m$ of the user are observed which serve as the \textit{context} for the decision. Then a decision is made which is encoded by the variable $p$. Finally, the user's response $y$ is observed as a result of the decision made. We emphasize that our approach can handle categorical, ordinal, and continuous data with respect to $x$, $p$, and $y$. As examples of these components, for the choice prediction problem, the contextual variables $x$ consist of features about the user (e.g., prior purchase history), the decisions $p$ correspond to the options offered by the firm to the user (e.g., an assortment of products), and the response $y$ indicates which option the user chose. For the bid landscape forecasting problem, the contextual variables $x$ encode the features describing the current user and auctioned ad spot (e.g., the ad spot's width/height), the decision $p \geq 0$ is the firm's submitted bid price, and the response $y \in \{0,1\}$  indicates the outcome of the auction  (win/loss). 

Our objective is to build an interpretable model for personalized decision-making problems that accomplishes two goals:
\begin{enumerate}
	\item \textbf{Market Segmentation}. Our model should yield a market segmentation of the contextual variables $x$. Here, we define a market segmentation as a partition of the context space $\mathcal{X}$ into a finite number of disjoint segments. 
	
	\item \textbf{Response Modeling}. Every market segment will correspond to a response model that we will estimate from that market segment's data.  Our response models should accurately estimate $\mathbb{P}(y|x,p)$, i.e., the probability of each response $y$ for all contexts $x$ and decisions $p$.  
\end{enumerate}

Section \ref{sec:model} presents our MST approach which tackles these tasks \textit{jointly}, with the market segmentation being informed by the resultant response models, \edit{and leads to an interpretable model for personalized decision-making problems.} 
Section \ref{sec:training} describes an algorithm for training MSTs from historical data.



\subsection{Market Segmentation Trees (MSTs)} \label{sec:model}

We tackle the personalized decision-making problem using an approach we call \textit{Market Segmentation Trees} (MSTs). MSTs perform market segmentation according to successive decision tree splits on the contextual variables $x$. Each split partitions the space of contexts with respect to a single contextual variable; continuous and ordinal contexts are split using inequalities (e.g., ``Age $\leq$ 40?''), while categorical contexts are split using equalities (e.g., ``Gender = Male?''). Each resulting market segment $l$ -- corresponding to  a \textit{leaf} of the MST and defined solely by contextual variables $x$ -- contains a response model $f_l(y|p)$ estimating the distribution of the response $y$ given the decision $p$ for users in segment $l$. Since different market segments may exhibit different distributions of the response $y$, the response models $f_l(y|p)$ may vary significantly across segments.

\begin{figure}[btp!]
\FIGURE{\centering
	\includegraphics[width=0.73\linewidth]{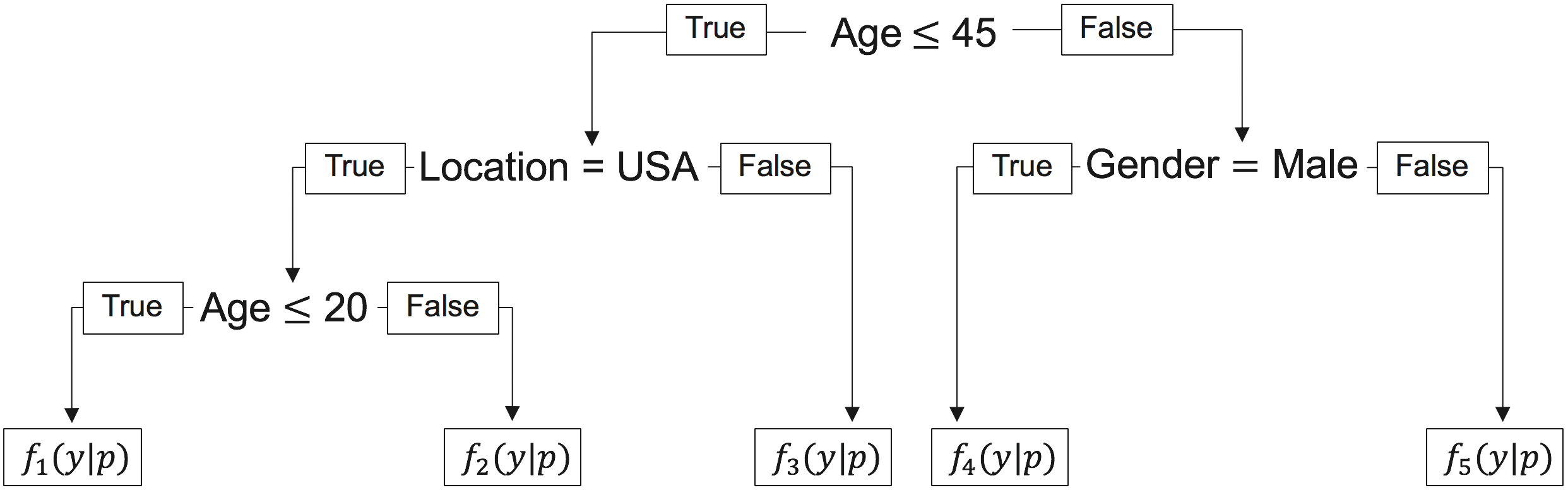}\label{fig:mt}
}{An example of an MST with five market segments.}
{Decision tree splits are performed with respect to the contextual variables \textit{age} (numeric), \textit{location} (categorical), and \textit{gender} (categorical).  Each of the resulting market segments contains a unique model $f_l(y|p)$ of the distribution of the response given the decision variables.} 
\end{figure}

\edit{The successive splits over contextual variables make MSTs interpretable and easy to use.} In a new prediction task, i.e. to estimate $\mathbb{P}(y|x,p)$ for a given context $x$ and decision $p$, one simply needs to follow the decision tree splits to the leaf $l$ to which the context $x$ belongs and output $f_l(y|p)$. For example, with respect to the MST in Figure \ref{fig:mt}, a user with context $x = \{$Age = 30, Location = USA, Gender = Male$\}$ would belong to segment $l =$ 2, so response model $f_2(y|p)$ would be used to make predictions with respect to that user's response behavior. As Figure \ref{fig:mt} demonstrates, the market segmentation produced by an MST may be easily visualized. Even in high-dimensional settings that may be too large to visualize, \edit{we consider the market segments to be interpretable since given a context $x$, one can follow a sequence of binary splits to describe its market segment and identify its corresponding response model.}
Since the \edit{most important} contextual variables \edit{with respect to differentiating response behavior} are already accounted for in the MST's decision tree splits, the response models can focus solely on the relationship between the decision variables and responses. \edit{This allows them to be simple  and leads to identical probabilistic responses for all observations within each market segment, facilitating their use and analysis for behavioral insights  across segments. We note that in situations where these characteristics of response models are not relevant, the response models in our MSTs could easily incorporate contextual features; we explore this idea further in our case study in Section \ref{sec:Swissmetro}.} 

MSTs also have a number of desirable properties as estimators. The decision tree splitting procedure is non-parametric, allowing MSTs to model potentially  non-linear relationships in the mapping from contexts to segments. MSTs also naturally model interactions among the contextual variables; for example, in the MST in Figure \ref{fig:mt}, the variable \textit{age} interacts with both \textit{location} and \textit{gender}.
\edit{We note that if interpretability and market segmentation are not required - as may be the case in some applications - a single, high-dimensional response model with interaction terms between contexts and decision variables may suffice, although this approach may still be computationally challenging and may require to specify a large number of interaction terms.}

MSTs provide a general framework that can be utilized to design new algorithms for various personalized decision-making problems. To do so, the practitioner simply needs to specify a family of response models for the given problem at hand, as well as a loss function for training the response models (see Section \ref{sec:training}, where this notion is described in greater detail). To demonstrate the versatility of the MST framework, we design two specialized MST algorithms for fundamental personalized decision-making problems in the following subsections.

\subsubsection{Choice Model Trees (CMTs)} \label{sec:cmts}

We propose a specialized MST algorithm, \textit{Choice Model Trees (CMTs)}, which can be used to predict a user's choice amongst multiple options. The CMT algorithm segments users on the basis of available demographic information and activity history on the site (e.g., prior purchases). Within each segment, a Multinomial Logit (MNL) choice model is fit as the response model to predict the probability that users in that market segment choose each option; MNL models are widely used for modeling user choice behavior \citep{train2009discrete}. Let $p = \{p_h\}_{h \in [H]}$ denote the collection of feature vectors encoding an offered assortment of $H$ options, with $p_h \in \mathbb{R}^q$ representing the feature vector encoding option $h \in [H] \coloneqq \{1,...,H\}$.  If the options correspond to different products, for example, then the elements of $p_h$ might include the products' price, color, and brand. Let $y \in \{0,1,...,H\}$ denote the user's choice when being presented with the assortment $p$ -- specifically, let
	\begin{equation*}
		y = \begin{cases}
			h, \text{if the user chooses option $h \in [H]$,}\\
			0, \text{if the user does not choose any option.}
		\end{cases}
	\end{equation*}
	
	Each leaf $l$ of the CMT contains an MNL instance, $f_l(y|p)$, estimating the probability of each outcome $y$ given the features $p$ describing the assortment of options. Let $\beta_l \in \mathbb{R}^q$ denote the parameters of the MNL model in leaf $l$. Then, the random utility that a user belonging to leaf $l$ experiences by choosing option $h$ is modeled as
 	\begin{equation*}
	U_h = \beta_l^T p_h + \epsilon_h  \,,
 	\end{equation*}

   \noindent while choosing no option has utility $U_0 = \epsilon_0$, where $\{\epsilon_h\}_{h=0}^H$ are random (Gumbel-distributed) noise terms independently and identically distributed across options. The user is assumed to be utility-maximizing, choosing option $h$ over $h'$ if $U_h > U_{h'}$, and choosing no option if none of the utilities are greater than the reference utility $U_0$. Thus, the probability of observing each choice takes the following form:
	\begin{equation} \label{eqn:mnl}
		\begin{cases}
			f_l\left(y=h \mid p\right) = \dfrac{e^{\beta_l^T p_h}}{1+\sum_{h' \in [H]}e^{\beta_l^T p_{h'}}}, \forall \, h \in [H] \\[0.6cm]
			f_l\left(y=0 \mid p\right) = \dfrac{1}{1+\sum_{h' \in [H]}e^{\beta_l^T p_{h'}}}
		\end{cases}
	\end{equation}
	Note that the number of options in the assortment ($H$) is permitted to vary across users. Our work also accommodates a noteworthy alternate form of the MNL model which allows for \textit{option-specific} parameters $\beta_{l,h}$, in which the utility from option $h$ takes the form $U_h = \beta_{l,h}^T p_h + \epsilon_h$. \edit{Such a form is valuable when the per-unit impact on utility of option features $p$ varies across options; for example, the per-unit disutility of price may differ from a luxury brand option to a generic brand option.} 

\subsubsection{Isotonic Regression Trees (IRTs)}
We propose a specialized MST algorithm, \textit{Isotonic Regression Trees} (IRTs), which can be used to solve the bid landscape forecasting problem. The tree segments ad spot auctions according to contexts such as the auctioned ad spot's dimensions in pixels and the visiting user's location. Here, an ad spot auction refers to the selling mechanism of a particular advertisement opportunity (e.g., location on website) for a particular user (e.g., visitor to website). Thus the ``market'' to be segmented in this application includes all instances of advertisement opportunities for users. Within each leaf of the tree, an isotonic regression model is used as the response model to estimate the bid landscape of the auctions belonging to that leaf. Let $p \geq 0$ denote an auction bid, and let $y$ be a binary variable which equals 1 if the bid won the auction and 0 if the bid lost. The isotonic regression model in each leaf $l$, denoted by $f_l(y|p)$, estimates the probability that a given bid $p$ will result in an outcome $y$ for auctions in that leaf. 

\begin{figure}[btp!]
\FIGURE
  {
	\centering
	\begin{subfigure}{.45\linewidth}
		\centering
		\includegraphics[width=0.88\linewidth]{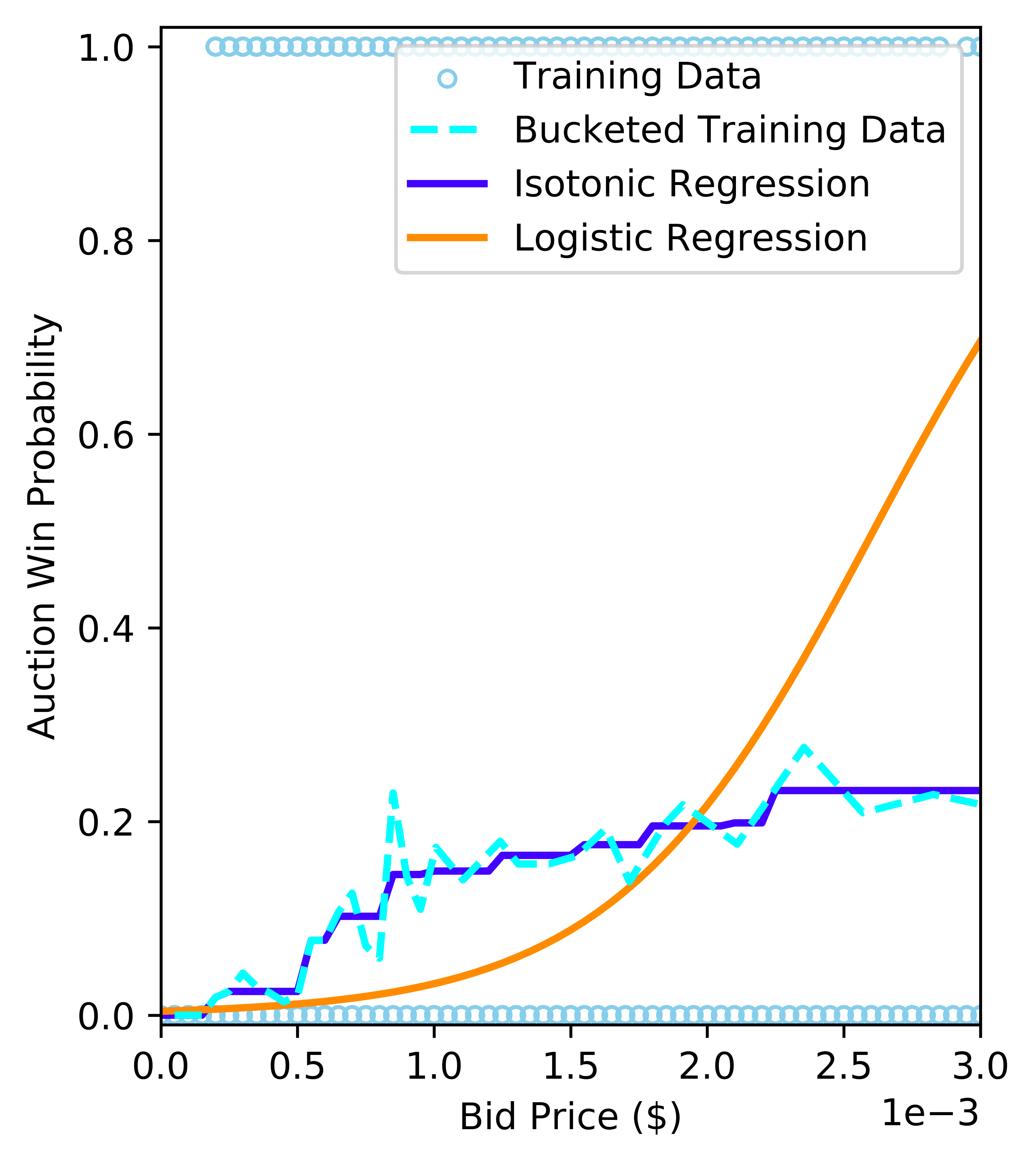}
		\caption{\small IR model (first leaf)}
		\vspace{0.1cm}
		\label{fig:sub1}
	\end{subfigure}%
	\begin{subfigure}{.45\linewidth}
		\centering
		\includegraphics[width=0.88\linewidth]{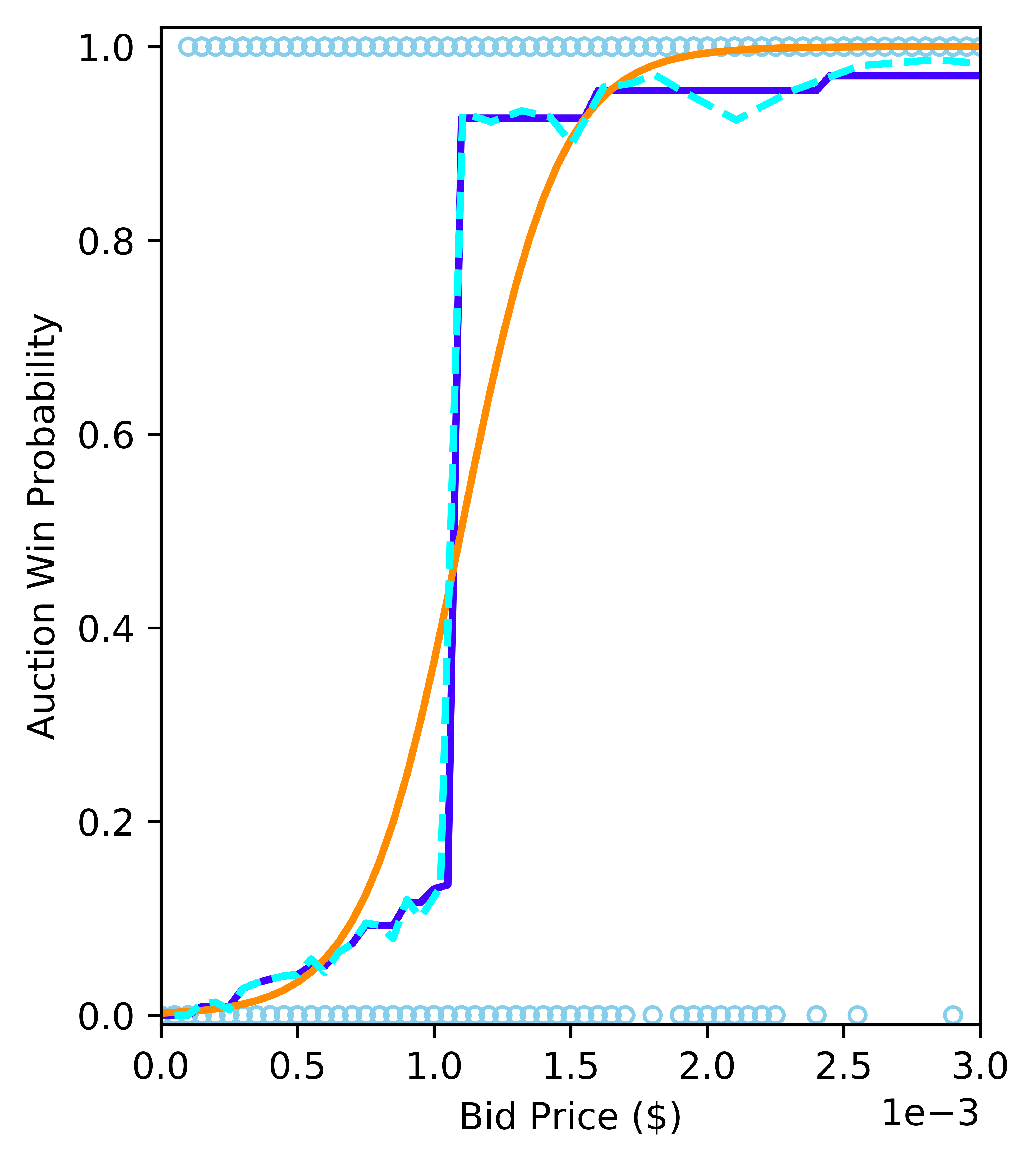}
		\caption{\small IR model (second leaf)}
		\vspace{0.1cm}
		\label{fig:sub2}
	\end{subfigure}
	}
{Estimated bid landscapes in two leaves of an IRT fit on bid data collected by a large DSP. \label{fig:ir}}
{The isotonic regression models are fit on training sets of auction outcomes (blue circles) within each leaf. Also included in the figures are logistic regression models trained on the same data. The models are compared against a curve (blue dashed line) constructed by bucketing the training set bids and computing the fraction of auction wins in each bucket.}
\end{figure}

An isotonic regression model is a free-form curve fitted to historical data in the following way: the curve is the best \textit{monotonically-increasing} curve that minimizes the training set mean squared error. The constraint of monotonicity is natural for this application, as the probability of an auction win should increase when the submitted bid $p$ increases. Isotonic regression models are non-parametric and uniformly consistent estimators, feasibly capturing any noisy, monotone function given sufficient data \citep{brunk1970estimation,hanson1973consistency}. To our knowledge, our paper is the first to propose their use for bid landscape forecasting. Also, the decision tree segmentation procedure of MSTs is non-parametric, imposing no distributional assumptions about the data. Thus, IRTs offer a fully non-parametric, interpretable algorithm for bid landscape estimation. 
Figure \ref{fig:ir} plots the estimated isotonic regression models in two different leaves of an IRT trained on historical bidding data collected by an anonymous DSP. As these plots demonstrate, different types of auctions can have differently-shaped bid landscapes, and isotonic regression models are flexible enough to capture these differences, whereas more commonly used parametric models like logistic regression are not. 

Beyond its application to bid landscape forecasting, IRTs also offer a powerful new tool for \textit{personalized pricing} \citep{elmachtoub2018value}. In these settings, the contextual variables $x$ are features encoding the visiting customer, the decision $p$ is the price of the offered product, and the response $y$ is a binary indicator of whether the customer purchased the product at that price. IRTs offer a non-parametric alternative for demand modeling which (1) naturally captures the monotonic (decreasing) relationship between product price and customer purchase probability through isotonic regression, and (2) finds a market segmentation driven by differences in customers' demand models.


\subsection{Training Procedure} \label{sec:training}
We present an algorithm for training the MSTs outlined in Section \ref{sec:model}. Let $n$ be the number of training set observations, and denote the collection of all such observations by $[n] =\{1,\ldots,n\}$. Let $i \in [n]$ denote an individual observation which consists of a context $x_i$, decision $p_i$, and response $y_i$. The training algorithm is fed the data $\{(x_i, p_i, y_i)\}_{i \in [n]}$ and learns (1) a segmentation of the contextual features $x_i$, and (2) the response models $f_l(y|p)$ within each segment. In Section \ref{sec:leafmods}, we first tackle problem (2) in isolation, showing how the final response models are fitted so that they accurately estimate the distributions of responses conditional on the decisions made.  We then propose in Section \ref{sec:segmentation} a training procedure for learning the market segmentation, which is driven by optimizing the accuracy of the resulting collection of response models. In Section \ref{sec:codebase}, we discuss our open-source code base for training MSTs. 
In Section \ref{sec:compcomplex}, we analyze the asymptotic computational complexity of MST training in terms of the depth of the tree and number of contextual variables. 

\subsubsection{Learning the Response Models} \label{sec:leafmods}

In what follows, we denote by $S_l \subseteq [n]$ the subset of training set observations which belong to leaf $l$ of the MST, and we designate by $f_l(y|p)$ the corresponding response model. Given a class $\mathcal{F}$ of response models, the goal is to find the response model $f_l \in \mathcal{F}$ which most accurately models the data $\{(p_i, y_i)\}_{i \in S_l}$. 
Specifically, our notion of model accuracy is captured by a loss function $\ell(p_i, y_i; f_l)$ which penalizes discrepancies between the observed response $y_i$ and the predicted response distribution $f_l(\cdot |p_i)$. We assume that this loss function is \textit{additive}, i.e., the loss incurred on the entire training data should be interpreted as the sum of the prediction losses for each individual observation. Each response model is trained by solving the following empirical risk minimization problem:
\begin{equation} \label{eqn:leafmodtrain}
\mathcal{L}(S_l) \coloneqq \min\limits_{f_l \in \mathcal{F}} \sum_{i \in S_l} \ell(p_i, y_i; f_l)
\end{equation}

To tailor our MST training algorithm to specific applications, the practitioner simply needs to specify a class of response models $\mathcal{F}$ and a loss function $\ell(p_i, y_i; f_l)$ for evaluating models $f_l \in \mathcal{F}$. Below are examples for how these are defined for the CMT and IRT models:

\begin{itemize}
    \item \textit{CMT}: The class of response models $\mathcal{F}$ are the set of MNL choice models  characterized by coefficients $\beta \in \mathbb{R}^q$ that satisfy Eq. (\ref{eqn:mnl}). MNL models are typically trained using the loss function of negative log-likelihood, defined as $\ell(p_i, y_i; f_l) := - \log(f_l(y=y_i|p_i))$.
    \item \textit{IRT}: Since the response $y_i$ is binary, then without loss of generality we may identify $\mathcal{F}$ as a class of functions $f_l(p)$ estimating the probability of $y=1$ given the user belongs to segment (leaf) $l$. Isotonic regression fits a monotonically increasing function to the training data which minimizes mean squared error. Consequently, we define $\mathcal{F}$ as the set of all monotonically-increasing functions $f_l: \mathbb{R} \rightarrow [0,1]$, and the loss function is defined as $\ell(p_i, y_i; f_l) := \big(y_i - f_l(p_i)\big)^2$.
\end{itemize}


\subsubsection{Learning the Segmentation} \label{sec:segmentation}

We now describe our market segmentation algorithm. From Eq. (\ref{eqn:leafmodtrain}), $\mathcal{L}(S_l)$ represents the total loss after training a response model on the collection of observations $S_l$. The goal of our market segmentation algorithm is to find the MST which segments the data into $L$ leaves, $S_1,...,S_L$, whose response models collectively minimize training set loss:
\begin{equation} \label{eqn:opt}
\min_{(S_1,...,S_L) \in {\mathcal P}(n)} 
\sum_{l=1}^L \mathcal{L}(S_l) \ ,
\end{equation}
\noindent where ${\mathcal P}(n)$ is the collection of decision tree  partitions satisfying $\bigsqcup_{l} S_l = [n]$.

It is clear that this optimization problem is NP-complete, since training optimal classification trees is a special case which is known to be NP-complete (to formulate a classification tree as an MST, let each response model map to a constant $K \in \{0,1\}$ and define the loss function as classification loss) \citep{laurent1976constructing}. Thus, we rely on a technique known as \textit{recursive partitioning}  to approximate an optimal market segmentation. The procedure is directly analogous to the CART algorithm for greedily training classification trees, recursively finding the best decision-tree split with the smallest loss across the resulting leaves \citep{breiman2017classification}. 

Denote the $j^{th}$ attribute of the $i^{th}$ observation by $x_{i,j}$. Starting with all of the data \edit{(i.e., $S_l = [n]$ for a single node $l = 1$)}, consider a decision tree split $(j,s)$ encoded by a splitting variable $j$ and split point $s$ which partitions the data into two leaves:
\begin{equation*}
\edit{S_{l,1}(j,s)} = \{i \in \edit{S_l} \mid x_{i,j} \leq s \} \,\,\, \text{and} \,\,\, \edit{S_{l,2}}(j,s) = \{i \in \edit{S_l}  \mid  x_{i,j} > s \} \,,
\end{equation*}
\noindent if variable $j$ is numeric, or
\begin{equation*}
\edit{S_{l,1}(j,s)} = \{i \in \edit{S_l}  \mid  x_{i,j} = s \} \,\,\, \text{and} \,\,\, \edit{S_{l,2}(j,s)} = \{i \in \edit{S_l}  \mid  x_{i,j} \neq s \} \,,
\end{equation*}

\noindent if variable $j$ is categorical. We wish to find the decision tree split $(j,s)$ resulting in the minimal loss incurred in leaves \edit{$S_{l,1}(j,s)$} and \edit{$S_{l,2}(j,s)$}, which corresponds to the following optimization problem:
\begin{equation} \label{eqn:greedy}
\min\limits_{j,s} \mathcal{L}(\edit{S_{l,1}(j,s)}) + \mathcal{L}(\edit{S_{l,2}(j,s)}) 
\end{equation}

This problem can be solved through an exhaustive search over all potential splitting variables and split points, choosing the split $(j,s)$ which achieves the best value of the objective function. When evaluating each split $(j,s)$, the data is partitioned according to the split and a response model is fit in each partition through solving Eq. (\ref{eqn:leafmodtrain}); the training errors from these models are then summed together to compute  objective function (\ref{eqn:greedy}). For continuous numerical variables, a search over all possible split points may be computationally infeasible, so instead the following approximation is used. The values of the continuous variable observed in the training data are sorted, and every $q^{th}$ quantile is evaluated as a candidate split point, where $q$ is a parameter chosen by the practitioner. In our numerical experiments, the value of $q$ varies between 2 and 10 depending on the application.

After a split is selected in this manner, the procedure is then recursively applied in the resulting leaves until a stopping criteria is met. Examples of stopping criteria include a maximum tree depth limit or a minimum number of training set observations per leaf. \edit{Pseudocode for our training procedure is provided in Algorithm \ref{alg}.} To prevent overfitting, the CART pruning technique detailed in \citet{breiman2017classification} can be applied to the MST after the training procedure using a held-out validation set of data. 

\begin{algorithm}[ht]
\edit{
\SetAlgoLined
\textbf{Input parameters}: Class of response models $\mathcal{F}$, loss function $\ell$\;
Boolean test $CHECK\_STOPPING\_CRITERIA(l,j,s)$ outputting $TRUE$ iff tree stopping criteria is not violated from splitting node $l$ according to context $x_{:,j}$ at split point $s$ (where $x_{:,j}$ is the vector of values for context $j$ across all training set observations)\;
\textbf{Initialize}: Internal nodes $\mathcal{T}^{Int} = \emptyset$, Leaf nodes $\mathcal{T}^{Leaf} = \emptyset$, Training nodes $\mathcal{T}^{Train} = \{1\}$\; Training dataset for root node $S_1 = [n]$\;
\While {$\mathcal{T}^{Train} \neq \emptyset$}
 {
 \For {$l$ $\in$ $\mathcal{T}^{Train}$}
 {\label{parallel}
 $\mathcal{C}_l = \{(j,s): CHECK\_STOPPING\_CRITERIA(l,j,s) = TRUE\}$\;
 \eIf{ $\mathcal{C}_l = \emptyset$}{
    $\mathcal{T}^{Leaf} = \mathcal{T}^{Leaf} \cup \{l\}$\;
    Fit response model $f_l$ to decision/response pairs $\{(p_i, y_i) : i \in S_l\}$ according to Eq. \eqref{eqn:leafmodtrain}\;
    $\mathcal{T}^{Train} = \mathcal{T}^{Train} \setminus \{l\}$\;
 }{
    $\mathcal{T}^{Int} = \mathcal{T}^{Int} \cup \{l\}$\;
    \For {$(j,s)$ $\in$ $\mathcal{C}_l$}
    {
        Partition training data $S_{l}$ into  segments $S_{l,1}(j,s), S_{l,2}(j,s)$ by splitting on context $x_{:,j}$ at split point $s$\;
        Fit response model to decision/response pairs $\{(p_i, y_i) : i \in S_{l,1}(j,s)\}$ according to Eq. \eqref{eqn:leafmodtrain}, compute loss $\mathcal{L}(S_{l,1}(j,s))$\;
        Fit response model to decision/response pairs $\{(p_i, y_i) : i \in S_{l,2}(j,s)\}$ according to Eq. \eqref{eqn:leafmodtrain}, compute loss $\mathcal{L}(S_{l,2}(j,s))$\;
    }
    Determine optimal split $(j_l^*, s_l^*) = \arg\min_{(j,s) \in \mathcal{C}_l} \mathcal{L}(S_{l,1}(j,s)) + \mathcal{L}(S_{l,2}(j,s))$\;
    Compute training dataset for left segment $S_{2l} = S_{l,1}(j_l^*, s_l^*)$\;
    Compute training dataset for right segment $S_{2l+1} = S_{l,2}(j_l^*, s_l^*)$\;
    $\mathcal{T}^{Train} = \mathcal{T}^{Train} \setminus \{l\} \cup \{2l, 2l+1\}$\;
 }
  }
  }
\textbf{Return}: Market segmentation $(j_l^*, s_l^*) \, \forall l \in \mathcal{T}^{Int}$, Response models $f_l \, \forall l \in \mathcal{T}^{Leaf}$  
 \caption{\label{alg} Market Segmentation Tree Training Procedure}
 }
\end{algorithm}

\subsubsection{Code Base for Training MSTs} \label{sec:codebase}

We provide an open-source implementation of our training procedure in Python \footnote{\url{https://github.com/rtm2130/MST}}. The implementation is general, allowing practitioners to specify the class of response models $\mathcal{F}$, loss function $\ell(p_i, y_i; f_l)$, and response model training procedure (i.e., procedure for solving Eq. (\ref{eqn:leafmodtrain})) which is best suited for their application. The stopping criteria used in training the MST is customizable as well: optional criteria include a maximum tree depth limit and a minimum number of observations per leaf. 

Our code offers several features for improved scalability on high-dimensional datasets. First, we develop a {\em parallelization scheme} to be used by our algorithm in the event that multiple processor cores are available. The main computational bottleneck of the training algorithm is in repeatedly solving the split selection optimization problem of Eq. (\ref{eqn:greedy}) to determine all internal splits of the MST. At a given depth of the MST, determining all splits at this depth can be thought of as independent subproblems which can be computed in parallel; thus, our parallelization strategy distributes all instances of the split selection optimization problem of Eq. (\ref{eqn:greedy}) at a given tree depth across any available processor cores. \edit{More specifically, our implementation parallelizes the `for loop' in line \ref{parallel} of Algorithm 1.} This parallelization scheme can significantly reduce or even nullify the effect of tree depth on computational complexity for a sufficiently large number of training observations (see Section \ref{sec:compcomplex}).

Second, we use the parameter estimates of the parent's response model as a {\em warm-start} to reduce the number of gradient descent iterations needed to fit the children's response models as part of the split selection optimization problem of Eq. (\ref{eqn:greedy}). Among all response models computed in the tree, parent nodes are arguably the most similar and informative estimates available. Moreover, this strategy evaluates and discards uninformative splits quickly, since in these cases the children's response model parameters are likely to be very similar to those of their parent and therefore training them requires very few iterations. Notably, we also apply a special warm-starting procedure when finding the optimal split point for a numerical variable. Any candidate split points for the numerical variable are evaluated in order of magnitude (e.g., ``$x < 1$'', then ``$x < 2$'', then ``$x < 3$'', etc.), and the response models corresponding to a particular split point are warm started with those from the previous split point. We find that the warm-starts significantly reduce the overall computational cost associated with learning the response models as part of the training procedure.

Finally, our code supports an adaptive optimization strategy to fit the response models. Namely, as the recursive partitioning training procedure progresses, the number of response models in the tree increases and the average number of observations per response model therefore decreases. We observe empirically that different stages of the training procedure may benefit from different response model optimization algorithms, adapted to the number of observations at hand. On large training sets common in the early splits in the tree, optimization algorithms that use mini-batching (e.g., stochastic gradient descent) may be required to efficiently fit the response models. However, as the recursion progresses and the tree depth increases, the computational burden shifts to fitting many small response models quickly, and thus, optimization methods with few gradient descent iterations like Newton's method are more efficient. Our code supports adapting the response model optimization algorithm used during the fitting process to the current number of observations. 

\subsubsection{Computational Complexity} \label{sec:compcomplex}

We provide theoretical bounds for the computational complexity of the MST training procedure as the number of training set observations becomes large. For ease of analysis, we assume throughout this section that the $m$ contextual variables are all binary and that the tree is trained to a fixed depth specified a priori by the practitioner. Let $n$ denote the number of training set observations, $m$ denote the number of contextual variables, and $D$ denote the depth of the MST being trained. We demonstrate two key properties, under some mild assumptions, of our training algorithm which illustrate its scalability to high-dimensional datasets:
\begin{enumerate}
    \item The computational complexity of our training algorithm is equivalent to fitting $O(D\cdot m)$ response models on training data of size $n$ (see Theorem \ref{thm:bound1}).
    \item Let $Q$ denote the number of cores available for parallel processing, and assume that the tree splits selected by the training algorithm are reasonably balanced. Then, the computational complexity of our training algorithm is equivalent to fitting $O(\max\{D/Q,1\}\cdot m)$ response models on training data of size $n$ (see Theorem \ref{thm:bound2}).
\end{enumerate}

Given that the number of response models in the MST scales exponentially in the tree's depth, one might expect the training algorithm's computational complexity to be exponential in $D$. However, we show through Theorem \ref{thm:bound1} that under reasonable technical assumptions, training time scales linearly in tree depth and in the number of contextual variables. Moreover, Theorem \ref{thm:bound2} implies that if the algorithm has access to a sufficiently large number of cores for parallel processing, i.e., if $Q$ is close in magnitude to $D$, then the effect of tree depth on training time can be greatly diminished or even nullified. Typically, compute nodes on high-performance computing clusters have at least 24 cores, and for many applications it is reasonable to expect MST depth to be less than 24.

{\blockedit
We now formally describe the framework and assumptions required to prove the above properties. 
Let $O(g(n))$ denote an upper bound on the computational cost of fitting a response model to training data of size $n$, i.e., the cost of solving the optimization problem in Eq. (\ref{eqn:leafmodtrain}). In other words, the cost of solving  Eq. (\ref{eqn:leafmodtrain})  is at most $g(n)$ times some fixed constant when $n$ is large enough. (Note that $g(n)$ can depend on the number of parameters being learned in the response model, but this is fixed and the same in every leaf. Thus, there is no need to capture this dependence explicitly.) For a given internal MST depth $d \leq D$, we number the nodes at depth $d$ according to $\{1,...,2^d\}$. Let $N_{\mathcal{T}}(d,l;n)$ denote the number of training set observations belonging to node $l \in \{1,...,2^d\}$ at depth $d$ of MST $\mathcal{T}$; note that $N_{\mathcal{T}}(D,l;n)$ is the number of observations belonging to each leaf $l$.

Our first theorem relies on the following technical assumptions (the formal definitions for any big-$O$ notation are provided in Appendices \ref{sec:proofthm1} and \ref{sec:proofthm2}):
\begin{assumption} \label{asn:gprop1}
$g(n)$ is continuous, increasing, and convex in $n$ for all $n \geq 0$, and $g(0)=0$. 
\end{assumption}
\begin{assumption} \label{asn:Nprop}
$N_{\mathcal{T}}(D,l;n) \rightarrow \infty$ as $n \rightarrow \infty$ for all $l=1,\ldots,2^D$.
\end{assumption}

As an example, note that Assumption  \ref{asn:gprop1} would be satisfied for ordinary linear regression by setting $g(n) = n $ (the complexity is linear in $n$). Assumption \ref{asn:Nprop} may be interpreted as a weak assumption on the distribution of the contextual variables in the training set. The assumption expresses that, for any partitioning of the contextual variables dictated by different MSTs of depth $D$, the number of observations in each leaf increases without bound as the sample size $n$ increases.
\begin{theorem} \label{thm:bound1}
If Assumptions  \ref{asn:gprop1} and \ref{asn:Nprop} hold, then the computational complexity of the MST's training algorithm may be expressed as $O\big(D\cdot m\cdot g(n)\big)$.
\end{theorem}

The proof of the theorem is presented in Appendix \ref{sec:proofthm1}. Theorem \ref{thm:bound1} implies that the complexity of the MST's training algorithm is equivalent to fitting $C_0 \cdot D\cdot m$ response models to the training data when $n$ is large enough, where $C_0$ is a constant independent of the problem parameters.  

Next, we analyze how the computational complexity of the training procedure is improved through use of the parallel processing scheme outlined in Section \ref{sec:codebase}. For depths $d = 0,1,...,D-1$, the training algorithm parallelizes the split selection procedure of Eq. (\ref{eqn:greedy}) across all nodes of depth $d$ within the MST. Note that all nodes across a given depth $d$ collectively partition the training set observations, i.e. $\sum_l N_{\mathcal{T}}(d,l;n) = n$. In order to effectively distribute each node's workload across the available cores for parallel processing, it is important that the partitioning of observations across nodes is not greatly imbalanced. Indeed, the worst case for parallel processing is for one node to contain all of the observations, in which case parallelization yields no benefits for our training algorithm. Thus, we assume that all splits chosen by the recursive partitioning procedure are reasonably balanced, i.e., they partition the data into roughly equal proportions:

\begin{assumption} \label{asn:equalprop}
Let $\mathcal{T}$ denote the trained MST. For all $d \in \{0,...,D-1\}$ and $l \in \{1,...,2^d\}$, $N_{\mathcal{T}}(d,l;n) = O(\frac{1}{2^d} \cdot n)$.
\end{assumption}
\begin{assumption} \label{asn:gprop2}
For any constant $C>0$, $g\big(Cn\big) = O\big(g(n)\big)$.
\end{assumption}  
To ensure Assumption \ref{asn:equalprop} holds in practice, one may restrict the split selection procedure of Eq. (\ref{eqn:greedy}) to only include splits which are not greatly imbalanced. This is arguably desirable from a learning perspective as well, as balanced splits can yield shallower and thus more interpretable decision trees.  Many runtime functions satisfy Assumption \ref{asn:gprop2}, including the complexity of computing the linear regression OLS estimator and, more generally, any function polynomial in $n$. For example, if $g(n) = n^a $, then $    g(Cn) = C^a n^a = C^a g(n) = O(g(n))$.

\begin{theorem} \label{thm:bound2}
If Assumptions   \ref{asn:gprop1}, \ref{asn:Nprop},   \ref{asn:equalprop}, and \ref{asn:gprop2} hold, then the runtime  of the MST's training algorithm with $Q$ parallel processors is $O\big(\max\{D/Q,1\}\cdot m \cdot g(n)\big)$.
\end{theorem}

Here we use the notion of runtime, which is equivalent to the time until all the $Q$ processors finish working \citep{jeje1992introduction}. Intuitively, Theorem \ref{thm:bound2} upper bounds the total computational cost by the maximum runtime on an individual processor times the number of processors.
}
The proof is given in Appendix \ref{sec:proofthm2}. Theorem \ref{thm:bound2} implies that the runtime of the training procedure is equivalent to fitting $O\big(\max\{D/Q,1\}\cdot m\big)$ response models to the training data. As discussed previously, we may diminish or even nullify the effect of tree depth on model complexity by setting $Q \approx D$, which is often feasible in practice.




\section{Numerical Experiments} \label{sec:experimental}

In this section, we evaluate the empirical performance of our MST methodology on several datasets. 
Our results demonstrate that MSTs are not only interpretable models but also yield competitive predictive performance of response behaviors when compared with other methods. 

\subsection{CMT Performance Evaluation}
\label{sec:CMTexperiments}

First, we apply the CMT algorithm to datasets derived from three ``ground truth'' models, each using a different method for simulating choice behavior. Second, we train and evaluate CMTs on \edit{a publicly available dataset describing transportation mode selection.}

\subsubsection{Experiments Using Synthetic Datasets} \label{sec:synthetic}

\paragraph{Dataset feneration.}  In each dataset, a user is encoded through four contextual variables ($x$) which can be used for the purposes of market segmentation. Each user is shown a random assortment ($p$) of 2-5 options, with each option encoded by four features (e.g., price). The user's response ($y$) to the assortment represents which option the user chose. The objective is to find a market segmentation of the contextual variables which leads to accurately predicting choice probabilities.

We generate 10 datasets -- including contexts, assortments, and choices -- from each of three different ``ground truth'' models, summarized below. Further details of how each dataset is generated are included in Appendix \ref{app:details-synthetic}. Each dataset is comprised of 25000 training set observations, 25000 validation set observations, and 25000 test set observations.

\begin{enumerate}
    \item ``Context-Free'' MNL: A single MNL model is used to simulate choices for all users. Contextual variables are simulated independently from choices, and therefore the contexts have no relevance to choice prediction. 
    
    \item Choice Model Tree: Choices are simulated through a Choice Model Tree of depth 3. The CMT maps users to leaves through decision tree splits on the users' contextual variables. Each leaf contains an MNL model used to simulate choices for all users belonging to that leaf. 
    
    \item $K$-Means Clustering Model: Choices are simulated according to the following procedure motivated by the popular $K$-means clustering market segmentation algorithm. Users belong to one of $K$ market segments, where $K$ is sampled from the possible values of $\{4,5,6,7\}$. Each segment $k \in \{1,...,K\}$ is associated with its own MNL model as well as a ``mean context vector'' $\bar{x}_k$. Each observation in the dataset is simulated by (1) sampling a market segment $k$ for the user, (2) sampling the user's context ($x$) from a multivariate normal distribution with mean parameter $\bar{x}_k$, and (3) sampling the user's choice ($y$) from segment $k$'s MNL model. 
\end{enumerate}

\paragraph{Experimental setup.} Using the training set observations for each of the generated datasets, CMTs are trained to depths of 0, 3, and 5, which correspond to 1, 8, and 32 leaves (i.e., market segments), respectively, and we prune the trees using the validation set observations according to the procedure described in \citet{breiman2017classification}. Note that the CMT of depth 0 is equivalent to a single, context-free MNL model, i.e., without market segmentation. We include CMTs of different depth sizes to examine the relationship between CMT model complexity and predictive accuracy. We also implement a $K$-means approach (MNL-KM) that uses training set observations to first perform $K$-means clustering on the contextual features ($x$) and then fit an MNL model within each cluster using only the decision variables ($p$). This clustering method represents a typical approach for market segmentation, whereby users are segmented based on feature dissimilarity rather than differences in their choice behavior. The number of clusters $K$ is tuned on a grid of values $\{1,2,...,K_{max}\}$ using the validation set observations. For each of the CMT depths we consider, we allow MNL-KM to utilize up to the same number of market segments as that CMT; for example, a CMT trained to a depth of 3 is compared against an MNL-KM utilizing at most $K_{max} = 2^3 = 8$ clusters. See \citet{friedman2001elements} for further background on $K$-means clustering.

Predictive accuracy on the test set observations is measured using {\em mean absolute error} (MAE), where \textit{absolute error} with respect to a single observation is defined as the average, taken over all options in the offered assortment, of absolute differences between each option's choice probability estimate and its \textit{true} choice probability specified by the ground truth model. To calculate MAE, we then average the absolute error over all observations in the test set. 


\paragraph{Results.}  We first evaluate the CMT and MNL-KM algorithms on 10 different datasets generated under the context-free MNL ground truth model in order to assess whether these approaches overfit on the contextual variables when they have no underlying relationship with the choice outcomes. 
As might be expected, we observe that the performance of the CMT and MNL-KM algorithms are equal when trained using a single market segment. Indeed, both a CMT of depth 0 and an MNL-KM with $K = 1$ equivalently represent a single context-free MNL model. Since the ground truth for these datasets is also a context-free MNL model, there is no model misspecification under either approach. Hence, the two algorithms achieve high levels of accuracy with average MAEs of less than 0.0025. 
We also observe that the CMT and MNL-KM algorithms achieve consistent test-set performance when permitted to utilize larger numbers of market segments. This signifies that the methodology used to prevent overfitting is working properly -- the CMT pruning algorithm always prunes the tree to depth 0 across the 10 datasets, and the MNL-KM algorithm always selects $K = 1$ through its tuning procedure.

\begin{figure}[btp!]
\FIGURE
  {
	\centering
	\begin{subfigure}{.48\linewidth}
		\centering
		\includegraphics[width=0.98\linewidth]{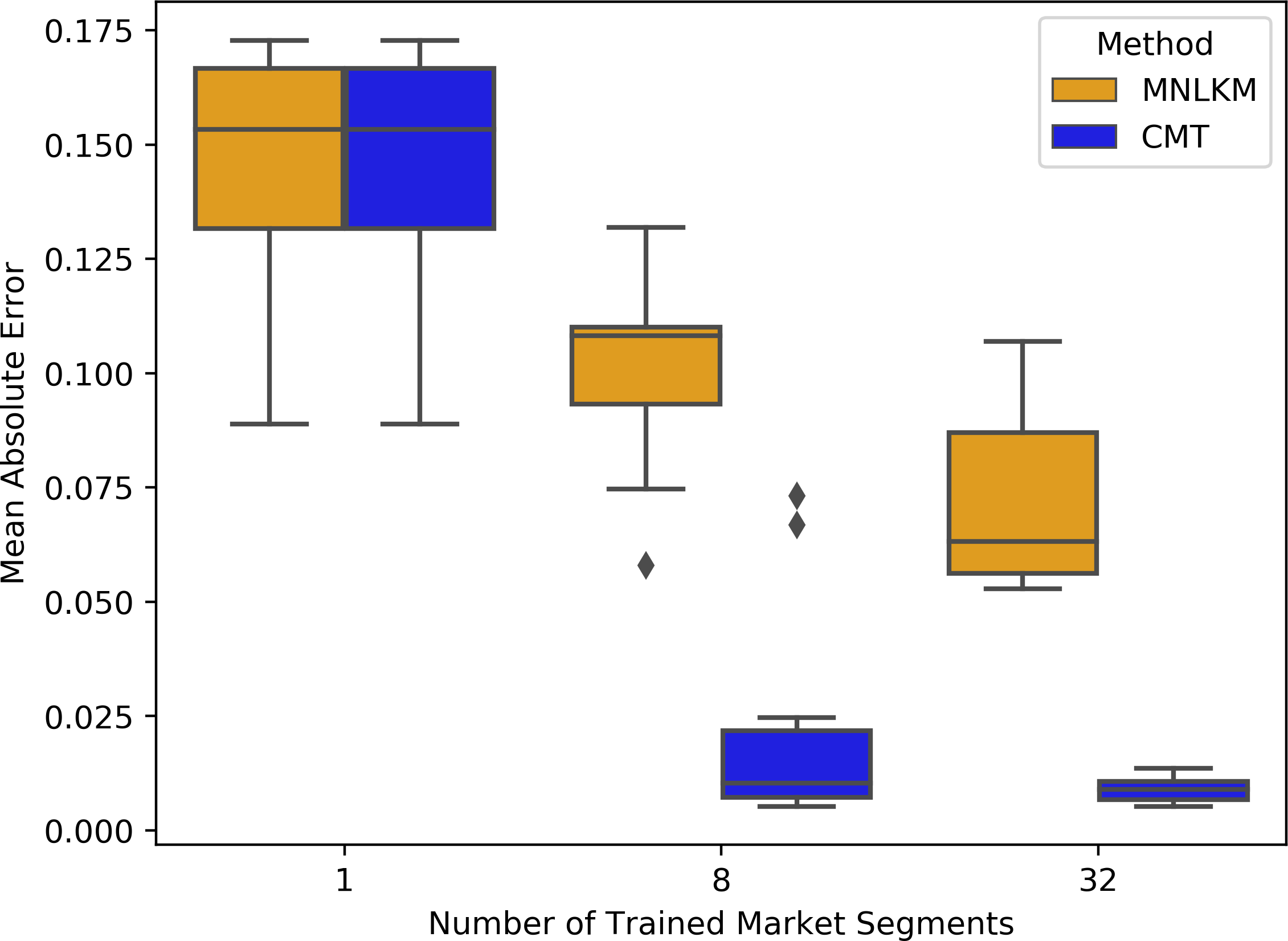}
		\caption{\footnotesize Test set MAEs incurred by the MNL-KM and CMT algorithms on the CMT ground truth.}
		\vspace{0.1cm}
		\label{fig:simboxcmt}
	\end{subfigure}%
	\begin{subfigure}{.48\linewidth}
		\centering
		\includegraphics[width=0.98\linewidth]{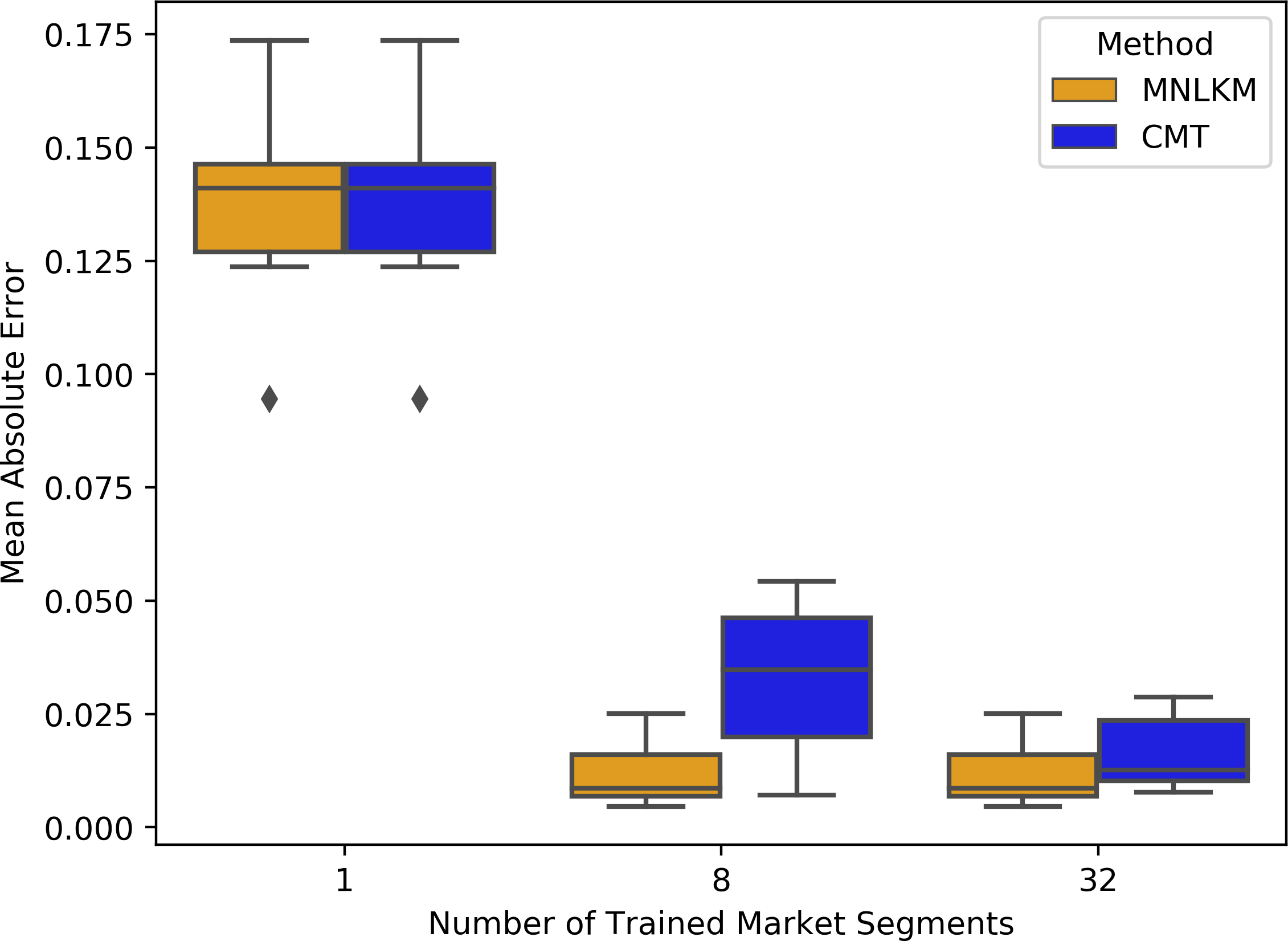}
		\caption{\footnotesize Test set MAEs incurred by the MNL-KM and CMT algorithms on the $K$-means clustering ground truth.}
		\vspace{0.1cm}
		\label{fig:simboxmnlkm}
	\end{subfigure}
	}
{Comparing CMT to MNL-KM with known ground truth. \label{fig:model_mis}}
{Each boxplot is constructed from the 10 datasets generated from the $K$-means clustering ground truth model.}
\end{figure}

We next evaluate the CMT and MNL-KM algorithms on 10 different datasets generated under the CMT ground truth model in order to assess whether CMTs are able to accurately recover the ground truth when presented with a sufficient number of training observations and to examine how MNL-KM performs under model misspecification. The prediction errors incurred by the algorithms on the test sets are visualized in Figure  \ref{fig:simboxcmt}. When the CMTs are trained to a depth of 3 (with 8 market segments), they often -- but not always -- recover the choice probability distributions. Recall that the CMT ground truths have a depth of at most 3, and some loss is incurred due to the suboptimality of the recursive partitioning heuristic. 
Nevertheless, we observe that when the CMTs are trained to a depth of 5, they are able to capture the choice probability distributions specified by the ground truth models almost perfectly. 
We also examine the performance of the MNL-KM algorithm on the same datasets. Although the market segmentations obtained by MNL-KMs improve prediction accuracy over the context-free models (i.e., $K = 1$), they fail to attain competitive performance relative to the CMT models. 
These findings demonstrate that MNL-KM is not necessarily robust to model misspecification. This is likely because MNL-KM does not consider the accuracy of the resulting collection of choice models when performing market segmentation; instead users are clustered solely on the basis of similarities in their contextual features.


We next evaluate the CMT and MNL-KM algorithms on 10 different datasets generated under the $K$-means clustering ground truth model. 
The prediction errors incurred by the algorithms on the test sets are visualized in Figure \ref{fig:simboxmnlkm}. We first observe that MNL-KM accurately recovers the response probability distributions specified by the ground truth model when the number of clusters $K$ is suitably large. However, we also observe that the CMT attains competitive predictive performance when trained to a suitably large depth of 5 (with the same number of leaves as $K$ used in MNL-KM). The CMT is therefore able to overcome the potential model misspecification introduced by the $K$-means clustering ground truth model. The CMT's robustness to model misspecification may be explained in part by its nonparametric decision tree splits, which permit the CMT to flexibly capture highly irregular mappings from contexts to market segments. Also, the CMT training algorithm is designed to yield a market segmentation which attains high choice prediction accuracy. 




{\color{black}
Finally, we believe that market segments defined by CMTs are typically more interpretable than those defined by the MNL-KM method, as we will describe at the end of Section \ref{sec:Swissmetro}.
}

\subsubsection{Experiment Using Swissmetro Dataset} \label{sec:Swissmetro}
{\color{black}
In this section, we train and evaluate CMTs on a publicly available dataset describing transportation mode selections for inter-city travel in Switzerland, known as the Swissmetro dataset \citep{bierlaire2001acceptance}. This dataset has been  used in previous research and software documentation to evaluate the fitting potential and predictability of various choice modeling methods, including MNL-based specifications~\citep{bierlaire2001acceptance} and neural networks~\citep{han2020neural}. 
The Swissmetro Stated Preference (SP) survey data was collected to gauge the demand for Swissmetro, a newly proposed underground system in Switzerland. 
Each respondent was asked to choose one mode of transportation out of a set of alternatives for inter-city travel. The choice set includes train (${\rm TRAIN}$), Swissmetro (${\rm SM}$), and car (${\rm CAR}$); the latter option was only available to car owners. Each mode of transportation is described via a number of option features, e.g., travel time and cost. In addition, information is collected about each individual, e.g., their gender and travel purpose. 
A detailed description of included features can be found in Appendix~\ref{app:details-swiss}.
We randomly split the data into a training set of 8041 observations as well as a validation set and a test set, each comprised of 1339 observations. This process is repeated ten times to generate 10 distinct train/validation/test sets. 

\paragraph{Experimental setup.} We evaluate the performance of our CMT algorithm with option-specific parameters $\beta_{l,h}$ compared to a variety of benchmarks. The primary set of benchmarks include a variety of ways to first segment customers by contextual features ($x$) and then fit an MNL model for each segment using only the decision variables ($p_h$):
\begin{itemize}
\item {\em MNL-KM:} We first perform $K$-means clustering to segment customers and then fit an MNL model within each segment.

\item {\em MNL-DT:} We first train a decision tree to segment customers and then fit an MNL model to each segment. The decision tree is built using the CART algorithm~\citep{breiman2017classification}; this supervised learning method recursively partitions the space of contexts $\{x_i\}_{i\in [n]}$ using the Gini criterion with respect to the chosen modes $\{y_i\}_{i\in [n]}$. 

\item {\em MNL-ICOT:} We first construct a decision tree to segment customers and then fit an MNL model to each segment. Here, we build the decision tree using the interpretable clustering approach of~\citep{bertsimas2020interpretable}, which is implemented via the Optimal Trees framework. Using a mixed-integer programming formulation, this unsupervised learning method creates clusters of contexts $\{x_i\}_{i\in [n]}$ that maximize an internal validation metric (e.g., Silhouette score); the cluster membership is represented by a tree-shaped partition, and the number of clusters is endogenously chosen.

\item {\em MNL:} This is a context-free MNL model using only decision variables ($p_h$); note that it corresponds to a CMT with depth zero and therefore consists of only a single market segment.

\end{itemize}

In Appendix \ref{app:AddCMT}, we also implement and evaluate two benchmarks that do \emph{not} provide a market segmentation in the form of a finite partition of the population. Instead, the response models use both contextual features ($x$) and decision variables ($p_h$) to predict users' responses.
To enable fair comparisons, we also implement modified versions CMT+, MNL-KM+, MNL-DT+, and MNL-ICOT+, where the MNL response models incorporate contextual information.

We performed our numerical experiments on an iMac Pro using 32000 MB of memory, parallelizing the CMT's training procedure across 8 processor cores. The CMT was trained using our open-source Python implementation with a minimum leaf size of 50 observations and a maximum trained depth size of 14. We specify the negative log-likelihood loss function from Section \ref{sec:leafmods} to score travel mode prediction error. For MNL-KM and MNL-DT, we call the $K$-means and CART methods of the scikit-learn library. Finally, the interpretable clustering method MNL-ICOT is executed using the ICOT repository. We use the training set to fit each method. Next, the hyper-parameters are tuned by minimizing the loss attained by the fitted models on the validation set. Specifically, the CMT is pruned using the mean squared error, according to the procedure described by \citet{breiman2017classification}. For MNL-KM, we tune the number of clusters in  $\{5 + 10\cdot k: k \in [30]\}$ using the mean squared error. For MNL-DT, we specify a maximum depth of 14 and prune the tree using the same criterion as CMT. For MNL-ICOT, since the running time exceeds 24 hours, an exhaustive search over hyper-parameters is computationally infeasible. Fortunately, this method does not require us to directly specify the number of clusters. We use the default parameters: \texttt{Geom\_search = True, threshold = 0.99, complexity\_c = 0.0, min\_bucket = 10, maxdepth = 10}. 

\begin{table}
\caption{Test set negative log-likelihoods (NLL) and mean squared errors (MSE) of the tested methods on 10 random splits of the data, labeled as S1 through S10.}
\label{tab:expresults}
	\begin{subtable}{1\linewidth}
	\vspace{0.2cm}
	\caption{Test set average NLL}
	\vspace{0.05cm}
	\resizebox{\textwidth}{!}{%
	    \setlength\tabcolsep{2pt}
		\begin{tabular}{@{}llllllllllllr@{}}
			\toprule
			Model & S1 & S2 & S3 & S4 & S5 & S6 & S7 & S8 & S9 & S10 & Avg. & \% Imp. \\ \midrule
CMT&{\bf 0.5875}&{\bf 0.6249}&{\bf 0.6350}&0.6547&{\bf 0.6227}&0.6564&{\bf 0.6194}&{\bf 0.5793}&0.6450&{\bf 0.5649}&{\bf 0.6190}& -\\
MNL&0.7873&0.8011&0.8108&0.8064&0.7904&0.7914&0.8024&0.7912&0.8149&0.7903&0.7986&22.5\%\\
MNL-KM&0.6301&0.6563&0.6413&0.6549&0.6424&{\bf 0.6290}&0.6462&0.6284&{\bf 0.5976}&0.6002&0.6326&2.1\%\\
MNL-DT&0.6295&0.6436&0.6422&{\bf 0.6473}& 0.6534&{ 0.6406}&0.6406&0.6119&0.6494&0.6116&0.6370&2.8\%\\
MNL-ICOT&0.6525&0.6611&0.6760&0.6786&0.6643&0.6709&0.6761&0.6407&0.6672&0.6421&0.6629&6.6\%\\
			 \bottomrule
		\end{tabular}
	}
	\end{subtable}

	\begin{subtable}{1\linewidth}
	\vspace{0.2cm}
	\caption{Test set MSE}
	\vspace{0.05cm}
	\resizebox{\textwidth}{!}{%
	    \setlength\tabcolsep{2pt}
		\begin{tabular}{@{}llllllllllllr@{}}
			\toprule
			Model & S1 & S2 & S3 & S4 & S5 & S6 & S7 & S8 & S9 & S10 & Avg. & \% Imp. \\ \midrule
CMT&{\bf 0.3499}&{\bf 0.3649}&{\bf 0.3750}&{\bf 0.3838}&{\bf 0.3670}&{\bf 0.3781}&{\bf 0.3641}&{\bf 0.3463}&0.3797&{\bf 0.3354}&{\bf 0.3644}&-\\
MNL&0.4884&0.4936&0.5091&0.5024&0.4904&0.4941&0.4953&0.4880&0.5075&0.4894&0.4958&26.5\%\\
MNL-KM&0.3822&0.3805&0.3841&0.3986&0.3856&0.3811&0.3922&0.3732&{\bf 0.3581}&0.3726&0.3808&4.3\%\\
MNL-DT&0.3817&0.3901&0.3930&0.3968&0.3982&0.3824&0.3869&0.3746&0.3952&0.3724&0.3871&5.8\%\\
MNL-ICOT&0.3951&0.4031&0.4179&0.4180&0.4048&0.4063&0.4117&0.3940&0.4058&0.3929&0.4050&10.0\%\\
			\bottomrule
		\end{tabular}%
	} \par
	\medskip
	{\footnotesize \textit{Note.} The column ``Avg.'' measures the average error across all 10 splits, and the column ``\% Imp.'' measures the percent improvement of CMT's average error compared to that of the benchmark.} 
	\end{subtable}
\end{table}

\paragraph{Results: Predictability.} Predictive accuracy on the test set observations is measured using the negative log-likelihood loss (NLL) described in Section~\ref{sec:leafmods} as well as mean squared error (MSE), which we define as follows. To avoid over-weighting low probability events in the calculation of NLLs, the choice probabilities are capped below at 1\% for all methods. The \textit{squared error} with respect to a single observation is defined as the sum, taken over all alternatives available, of squared differences between each travel mode probability estimate and its realized 0/1 choice outcome, i.e., for every observation $i \in [n]$, $\ell(p_{i,h},y_i, f_l) = \sum_{h \in H} (\mathbb{I}[y_i = h] - f_l(h|p_i))^2$. Mean squared error is then defined as the average squared error over all observations in the test set.  This metric is also referred to in the literature as the \textit{Brier score} and is a proper scoring rule for evaluating probabilistic predictions. 

The test set performance of the fitted CMTs and benchmarks that provide market segments are reported in Table \ref{tab:expresults} for each of the 10 random splits. CMT emerges as the most predictive choice model by a significant margin, followed by MNL-KM and MNL-DT. 
Appendix \ref{app:AddCMT} further illustrates the predictive power of our CMT approach when market segmentation is not important.

For a more fine-grained reading of these results, let us first recall that all tested methods have in common an underlying MNL probabilistic choice model, but they differ through how customer variables are accounted for. On both performance metrics, there is a substantial performance gap between the context-free MNL model and all other methods. Hence, these gains in predictive accuracy should be firstly attributed to the value of personalization, i.e., the ability to exploit customer-level features. Using Appendix \ref{app:AddCMT} to compare our metrics for MNL and MNL-INT reveals that interacting the context variables with decision variables cuts this performance gap approximately in half. Yet, the ability of CMTs to capture more complex non-linear dependencies between contexts and decisions yields a nearly equal marginal improvement of fitting quality. 

Of the benchmarks that conduct market segmentation, MNL-KM performs the best with respect to both NLL and MSE. 
Thus, the difference in performance between CMT and MNL-KM (on average, $2.1\%$ in NLL and $4.3\%$ in MSE) is instructive to measure the value-added of jointly segmenting the customer population and learning the response models. This difference is very explicit in Figure~\ref{fig:perf-func-K}, where we plot our accuracy metrics as a function of the number of market segments. According to our in-sample metrics, CMTs provide a much better fit to the data than the MNL-KM method; this is likely related to the fact that our recursive partitioning method considers the downstream accuracy of the fitted MNL models. The out-of-sample accuracy is also higher in most settings, albeit by a smaller margin. An important characteristic is that for a given value of NLL or MSE, the CMT algorithm typically results in far fewer market segments than the MNL-KM algorithm; for example, a CMT with only 50 market segments has approximately the same predictive power as the MNL-KM model with 100 market segments. 

The difference in performance of MNL-DT compared to CMT shows the limitations of using a standard decision tree heuristic to construct a segmentation driven by differences in customers' choices. This benchmark can be viewed as an ``ill-specified'' variant of our recursive partitioning method, whereby each split is evaluated based on the fit quality of a naive response model that overlooks the decision variables. MNL-ICOT seems to further handicap a decision tree approach by also not accounting for choice data when developing the tree.
 
 \begin{figure}[!htbp]
\centering
\caption{Performance of CMT and MNL-KM as a function of the number of market segments.}\label{fig:perf-func-K}
\begin{subfigure}{.5\linewidth}
\centering
\includegraphics[width=0.99\linewidth]{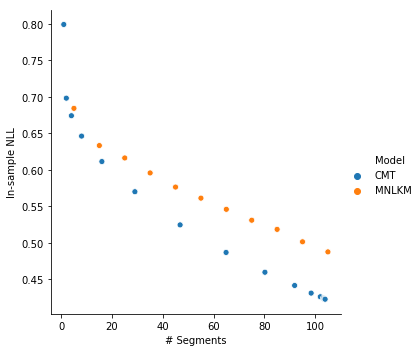}
\label{fig:nll1}
\end{subfigure}%
\begin{subfigure}{.5\linewidth}
\centering
\includegraphics[width=0.99\linewidth]{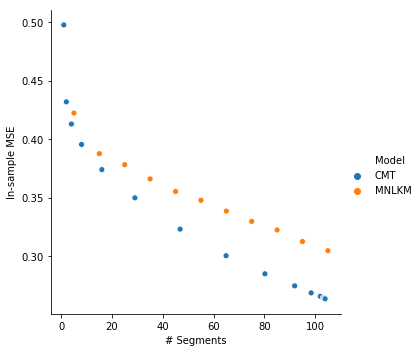}
\label{fig:mse1}
\end{subfigure}

\begin{subfigure}{.49\linewidth}
\centering
\includegraphics[width=0.99\linewidth]{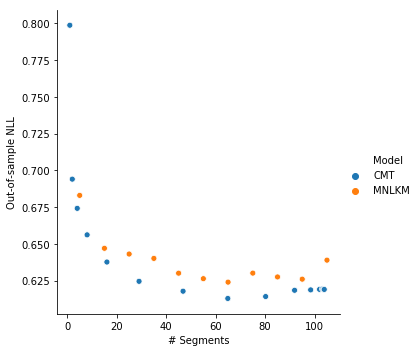}
\label{fig:nll2}
\end{subfigure}
\begin{subfigure}{.49\linewidth}
\centering
\includegraphics[width=0.99\linewidth]{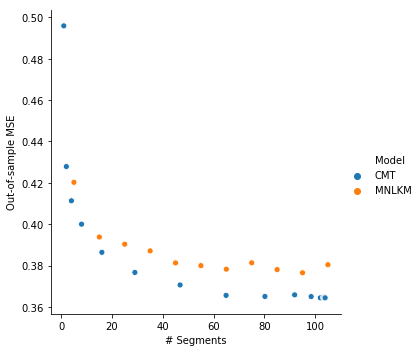}
\label{fig:mse2}
\end{subfigure}
\par
	\medskip
	{\small  \textit{Note.} The top two figures are the in-sample NLL and MSE of the CMT and MNL-KM methods, while the bottom two figures show the corresponding out-of-sample performances. For CMT, we vary the tree depth from 1 to 15. For MNL-KM, we vary the number of segments from 1 to 300.  In-sample metrics are computed without parameter tuning (number of clusters of MNL-KM and tree pruning of CMT). Out-of-sample metrics are computed after tuning the  parameters on the validation set.}
\end{figure}

\paragraph{Results: Interpretability.} 
As illustrated by Figure~\ref{fig:perf-func-K}, CMTs can provide a more parsimonious segmentation than the MNL-KM method without any loss of accuracy, which is desirable from an interpretability standpoint. 
Table~\ref{tab:intresults} further illustrates that parsimonious CMT models have better predictive accuracy than similarly parsimonious MNL-DT, MNL-KM, and MNL-ICOT models.
Namely, recall that the number of segments of MNL-ICOT is endogenously determined; specifically, fitting MNL-ICOT on our 10 random splits resulted in an identical tree formed by 11 market segments (tree depth of 4). Thus for each of the 10 random splits, we re-train MNL-KM with 11 market segments and re-train CMT and MNL-DT with a tree depth of 4 since the number of market segments cannot be pre-specified.
We note that at a tree depth of 4, the CMT and MNL-DT models can have a maximum of 16 market segments, more than the 11 market segments in the MNL-ICOT and MNL-KM models; thus we also re-train and compare the CMT and MNL-DT models with a tree depth of 3 (maximum of 8 market segments).
Table~\ref{tab:intresults} details the negative log-likelihoods and mean squared errors for each method across all 10 random splits. 
We see that when CMT, MNL-DT, and MNL-ICOT are constrained to produce equally interpretable tree-based market segmentations (tree depth 4), CMT has a significant advantage in predictive accuracy with respect to both NLL (3.8\% - 5.7\%) and MSE (4.6\% - 6.1\%).

\begin{table}
\caption{Comparing methods under a fixed number of market segments.}
\label{tab:intresults}
	\begin{subtable}{1\linewidth}
	\vspace{0.2cm}
	\caption{Test set average NLL}
	\vspace{0.05cm}
	\resizebox{\textwidth}{!}{%
	    \setlength\tabcolsep{2pt}
		\begin{tabular}{@{}lllllllllllllr@{}}
			\toprule
			Model & \#  & S1 & S2 & S3 & S4 & S5 & S6 & S7 & S8 & S9 & S10 & Avg. & \% Imp. \\ \midrule
CMT&16&{\bf 0.6293}&{\bf 0.6301}&{\bf 0.6510}&{\bf 0.6608}&{\bf 0.6316}&{\bf 0.6447}&{\bf 0.6368}&{\bf 0.6176}&{\bf 0.6539}&{\bf 0.6204}&{\bf 0.6376}& -\\
CMT&8&0.6503&0.6480&0.6627&0.6780&0.6596&0.6656&0.6597&0.6310&0.6670&0.6393&0.6561&2.8\%\\
MNL-KM&11&0.6315&0.6494&0.6800&0.6760&0.6603&0.6572&0.6701&0.6411&0.6542&0.6391&0.6559&2.8\%\\
MNL-DT&16&0.6705&0.6672&0.6939&0.6882&0.6694&0.6957&0.6793&0.6515&0.6830&0.6599&0.6759&5.7\%\\
MNL-DT&8&0.6705&0.6749&0.6939&0.6965&0.6694&0.6957&0.6937&0.6605&0.6917&0.6709&0.6818&6.5\%\\
MNL-ICOT&11&0.6525&0.6611&0.6760&0.6786&0.6643&0.6709&0.6761&0.6407&0.6672&0.6421&0.6629&3.8\%\\			
			 \bottomrule
		\end{tabular}
	}
	\end{subtable}

	\begin{subtable}{1\linewidth}
	\vspace{0.2cm}
	\caption{Test set MSE}
	\vspace{0.05cm}
	\resizebox{\textwidth}{!}{%
	    \setlength\tabcolsep{2pt}
		\begin{tabular}{@{}lllllllllllllr@{}}
			\toprule
			Model & \#  & S1 & S2 & S3 & S4 & S5 & S6 & S7 & S8 & S9 & S10 & Avg. & \% Imp. \\ \midrule
CMT&16&{\bf 0.3819}&{\bf 0.3819}&{\bf 0.3974}&{\bf 0.4073}&{\bf 0.3831}&{\bf 0.3876}&{\bf 0.3848}&{\bf 0.3736}&{\bf 0.3901}&{\bf 0.3761}&{\bf 0.3864}& -\\
CMT&8&0.3948&0.3955&0.4086&0.4214&0.4016&0.4027&0.4010&0.3824&0.4028&0.3891&0.4000&3.4\%\\
MNL-KM&11&0.3846&0.3893&0.4164&0.4159&0.4011&0.3956&0.4044&0.3954&0.3945&0.3907&0.3988&3.1\%\\
MNL-DT&16&0.4100&0.4035&0.4280&0.4239&0.4040&0.4203&0.4110&0.3990&0.4148&0.4021&0.4117&6.1\%\\
MNL-DT&8&0.4100&0.4101&0.4280&0.4341&0.4040&0.4203&0.4220&0.4051&0.4232&0.4127&0.4170&7.3\%\\
MNL-ICOT&11&0.3951&0.4031&0.4179&0.4180&0.4048&0.4063&0.4117&0.3940&0.4058&0.3929&0.4050&4.6\%\\			
			\bottomrule
		\end{tabular}%
	} \par
	\medskip
	{\footnotesize \textit{Note.} The column ``\#'' refers to the maximum number of market segments across  10 random splits. The column ``Avg.'' measures the average error across all 10 splits, labeled as S1 through S10. The column ``\% Imp.'' measures the percent improvement of CMT's average error compared to that of the benchmark.}
	\end{subtable}
\end{table}


}

{\color{black}
Compared to MNL-KM constrained by the number of market segments, CMT also achieves improvements of predictive accuracy: 2.8\% with respect to NLL and 3.1\% with respect to MSE. Although these improvements are smaller in magnitude than CMT's improvements over the tree-based methods, we believe that the CMT model with tree depth 4 leads to considerably more interpretable market segments than the MNL-KM method.
To see this, we note that each market segment from CMT can be described using a maximum of four features, whereas all features must be used to describe each market segment from MNL-KM. 
Relatedly, for CMT, identifying which segment an observation belongs to simply requires answering a sequence of at most four binary questions, each using only one feature; in contrast, for MNL-KM, identifying which segment an observation belongs to requires calculating the distance between the $m$-dimensional observation and the centroids of all segments, and then finding the closest segment. 
Because of this, if two very similar observations are assigned to different segments, it is clear from CMT which feature dictated that the observations were in different segments, whereas this is not the case for MNL-KM.
See Appendix \ref{app:CMTS10} for an example tree defining market segments produced by the CMT for split S10{\color{black}, as well as an example illustrating the interpretability of the tree and resultant outcome functions.}
}

\subsection{Isotonic Regression Tree Performance Evaluation} \label{sec:DSP}

We train and evaluate IRTs on bidding data from a Demand Side Platform (DSP), which will remain anonymous for confidentiality reasons. The DSP provided us with several weeks of bidding data across three different ad exchanges.
Each observation in the data is encoded by (1) the user and ad spot auction features available to the bidder ($x$), (2) the submitted bid price ($p$), and (3) the auction outcome (win/loss) ($y$). A detailed description of the user and ad spot auction features is included in Appendix \ref{app:details-DSP}.

\paragraph{Experimental setup.} For each ad exchange, an IRT is trained on a dataset of historical bids submitted by the DSP between 1/13/2019 and 1/24/2019, which amount to a training set of 60-370 million bids per exchange. The IRT is pruned using a validation set holding out 15\% of the training data following \citet{breiman2017classification}. Finally, the IRT is evaluated on test sets of bids submitted between 1/25/2019 and 1/31/2019 amounting to 40-160 million bids per exchange.
We compare the IRT algorithm's predictive performance with the following benchmarks; in selecting which benchmarks to use, we consider models which perform market segmentation and produce monotonically-increasing bid landscape curves. 

\begin{itemize}
	\item \textit{Const}: A model which predicts a constant win probability for all bid prices equal to the average training set win rate. 
	\item \textit{IR}: An isotonic regression model fit on the entire training set to estimate the auction win rate given the submitted bid price. This is a ``context-free'' model and does not incorporate features $x$.
	\item \textit{IRKM}: Performs $K$-means clustering on features $x$ and then fits an isotonic regression model within each cluster; the number of clusters $K$ is tuned using the validation set. 
	\item \textit{DSP}: The bid landscape forecasting model which the DSP used in production during the testing period, which was also trained using the same data as our training set. 
	\item \textit{LR, LRKM, LRT}: We include analogous benchmarks testing the impact of using logistic regression models as opposed to isotonic regression models. Logistic regression is one of the most common parametric approaches for probabilistically modeling binary response data. 
	The benchmark LR fits a single, ``context-free'' logistic regression model to the entire data; the benchmark LRKM performs $K$-means clustering on the auction features and fits a logistic regression model in each cluster; and the benchmark LRT runs our MST algorithm with logistic regression response models.
\end{itemize}

 
We conducted our experiments on a Dell PowerEdge M915 Linux server using 50000 MB of memory, parallelizing the IRT's training procedure across 8 processor cores. The IRT was trained on each exchange separately using our open-source Python implementation, specifying a minimum leaf size of 10000 observations and no depth limit. The IRT was trained and pruned using the mean-squared-error (MSE) metric, which measures the average squared difference between the algorithms' win probability estimates and the realized auction outcomes. The training procedure terminated after 12-35 hours of computational time across the three exchanges. Next, the trees were pruned on a validation set, taking 6-35 minutes to complete per exchange. {\color{black} The reasonable computation times of our training and pruning procedures illustrate the scalability of our implementation when presented with large-scale high-dimensional data. The final IRTs were of depths 52-78 and contained 800-4100 leaves. Although these performance-optimized IRTs are too large to be visualized, they may still be regarded as interpretable bid landscape forecasting models since they map each auction to a single response model (bid curve) that is solely a function of price; therefore the bid curve can be easily visualized and analyzed for bidding insights. At the end of this section, we also compare the performance of smaller, more interpretable IRTs against equivalently-sized IRKM benchmarks by plotting the prediction accuracy of both methods as a function of the number of segments.}



\begin{figure}
\centering
\caption{Test set ROC curves and AUCs of our algorithm (IRT) and the benchmarks on three ad exchanges.}\label{fig:roc}
\begin{subfigure}{.5\linewidth}
\centering
\includegraphics[width=0.99\linewidth]{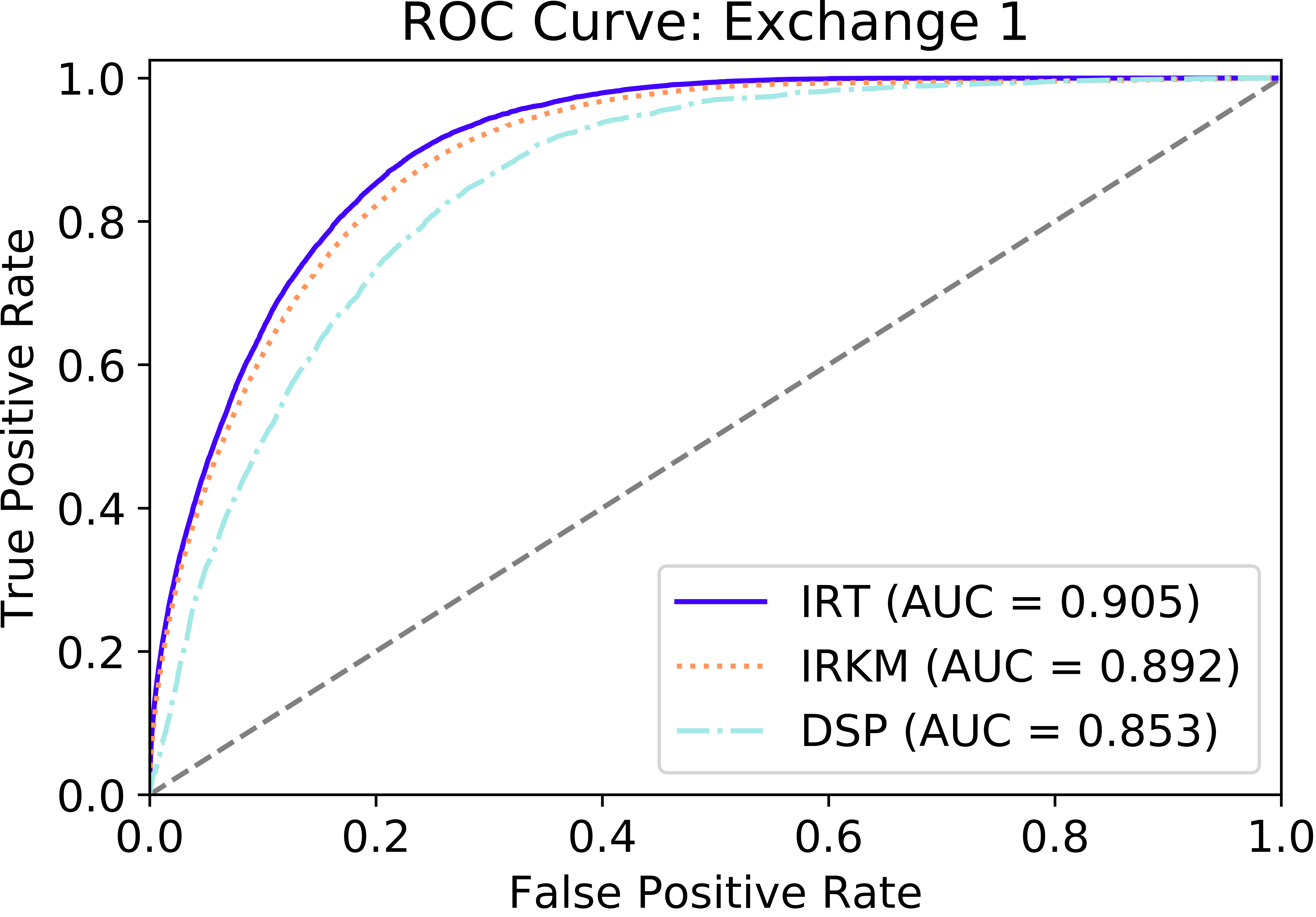}
\label{fig:roc1}
\end{subfigure}%
\begin{subfigure}{.5\linewidth}
\centering
\includegraphics[width=0.99\linewidth]{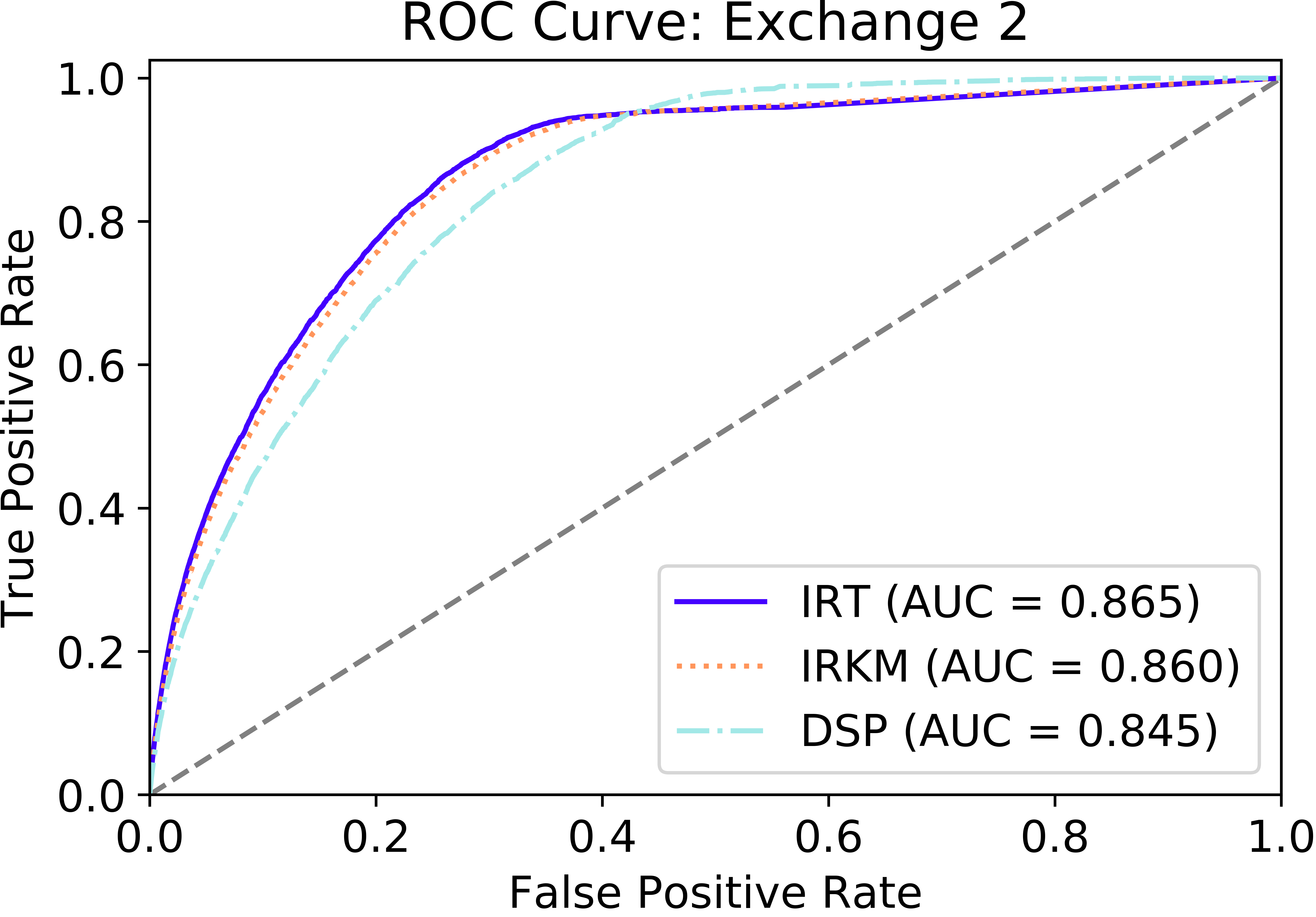}
\label{fig:roc2}
\end{subfigure}\\[1ex]
\begin{subfigure}{\linewidth}
\centering
\includegraphics[width=0.495\linewidth]{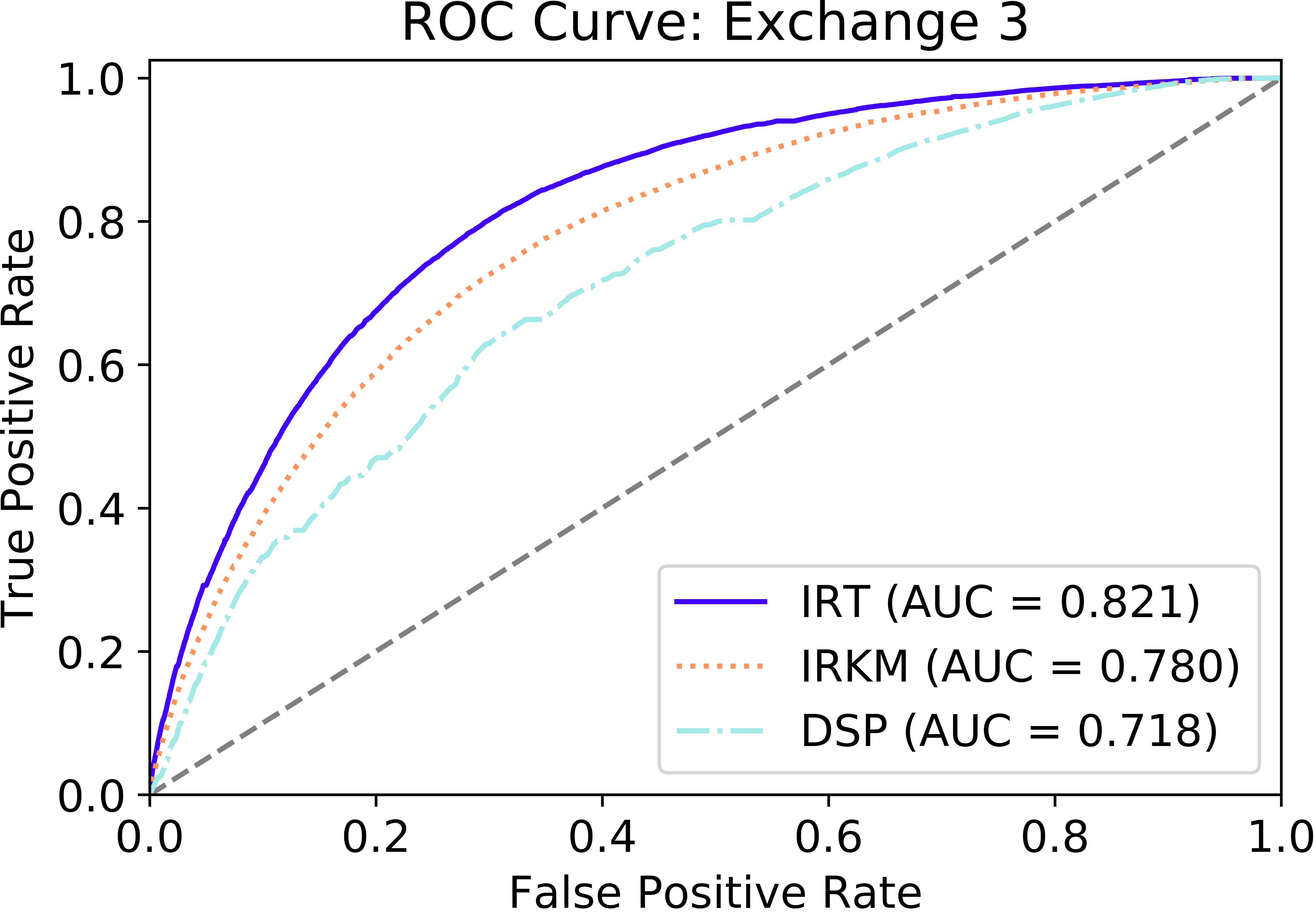}
\label{fig:roc3}
\end{subfigure}
\par
	\medskip
	{\footnotesize \textit{Note.} The benchmark IR, not shown in the figures due to space constraints, achieved AUCs of 0.844, 0.776, and 0.716 on exchanges 1,2, and 3, respectively.}
\end{figure}

\paragraph{Results: Predictability.} The test set performance of the IRT and benchmarks for each ad exchange is given in Table \ref{tab:mmresults}, in which we report (1) overall MSE measured across the entire test data, and (2) the MSEs for each individual day of test data. The algorithms were also compared on the basis of their test-set ROC curves using the AUC (area under curve) metric. The ROCs and AUCs obtained by the algorithms are described by Figure \ref{fig:roc}.
\begin{table}[htbp!]
\caption{Test set mean squared errors (MSEs) of our algorithm (IRT) and the benchmarks on three ad exchanges.}
\label{tab:mmresults}
	\begin{subtable}{1\linewidth}
	\vspace{0.2cm}
	\caption{Test set MSEs: Exchange 1}
	\vspace{0.05cm}
	    \setlength\tabcolsep{6pt}
		\begin{tabular}{@{}llllllllll@{}}
			\toprule
			Model & 1/25 & 1/26 & 1/27 & 1/28 & 1/29 & 1/30 & 1/31 & Avg. & \% Imp. \\ \midrule
			IRT & \textbf{0.0465} & \textbf{0.0476} & \textbf{0.0432} & \textbf{0.0474} & \textbf{0.0482} & \textbf{0.0539} & \textbf{0.0482} & \textbf{0.0480} \\
			LRT & 0.0508 & 0.0508 & 0.0458 & 0.0504 & 0.0523 & 0.0588 & 0.0521 & 0.0518 & 7.3\%\\
			Const & 0.0613 & 0.0613 & 0.0552 & 0.0599 & 0.0626 & 0.0718 & 0.0631 & 0.0625 & 23\% \\
			IR & 0.0538 & 0.0545 & 0.0492 & 0.0529 & 0.0540 & 0.0619 & 0.0550 & 0.0546 & 12\% \\
			LR & 0.0586 & 0.0584 & 0.0526 & 0.0571 & 0.0590 & 0.0680 & 0.0597 & 0.0593 & 19\% \\
			IRKM & 0.0489 & 0.0497 & 0.0446 & 0.0488 & 0.0494 & 0.0556 & 0.0497 & 0.0497 & 3.4\% \\
			LRKM & 0.0535 & 0.0540 & 0.0478 & 0.0522 & 0.0536 & 0.0603 & 0.0536 & 0.0537 & 11\% \\
			DSP & 0.0564 & 0.0558 & 0.0508 & 0.0560 & 0.0569 & 0.0640 & 0.0592 & 0.0572 & 16\% \\ \bottomrule
		\end{tabular}
	\end{subtable}

	\begin{subtable}{1\linewidth}
	\vspace{0.2cm}
	\caption{Test set MSEs: Exchange 2}
	\vspace{0.05cm}
	    \setlength\tabcolsep{6pt}
		\begin{tabular}{@{}llllllllll@{}}
			\toprule
			Model & 1/25 & 1/26 & 1/27 & 1/28 & 1/29 & 1/30 & 1/31 & Avg. & \% Imp. \\ \midrule
			IRT & \textbf{0.0276} & \textbf{0.0253} & \textbf{0.0341} & \textbf{0.0318} & \textbf{0.0366} & \textbf{0.0419} & \textbf{0.0405} & \textbf{0.0339} \\
			LRT & 0.0301 & 0.0273 & 0.0368 & 0.0344 & 0.0393 & 0.0450 & 0.0437 & 0.0366 & 7.3\%\\
			Const & 0.0316 & 0.0285 & 0.0391 & 0.0364 & 0.0414 & 0.0471 & 0.0451 & 0.0384 & 12\%\\
			IR & 0.0305 & 0.0275 & 0.0371 & 0.0349 & 0.0397 & 0.0449 & 0.0432 & 0.0368 & 7.9\%\\
			LR & 0.0320	& 0.0287 & 0.0394 & 0.0366 & 0.0417 & 0.0473 & 0.0455 & 0.0387 & 12\%\\
			IRKM & 0.0281 & 0.0258 & 0.0345 & 0.0321 & 0.0369 & 0.0423 & 0.0408 & 0.0343 & 1.2\%\\
			LRKM & 0.0306 & 0.0278 & 0.0372 & 0.0347 & 0.0396 & 0.0453 & 0.0440 & 0.0370 & 8.4\%\\
			DSP & 0.0296 & 0.0285 & 0.0377 & 0.0341 & 0.0379 & 0.0428 & 0.0416 & 0.0359 & 5.6\%\\ \bottomrule
		\end{tabular}%
	\end{subtable}

	\begin{subtable}{1\linewidth}
	\vspace{0.2cm}
	\caption{Test set MSEs: Exchange 3}
	\vspace{0.05cm}
	    \setlength\tabcolsep{6pt}
		\begin{tabular}{@{}llllllllll@{}}
			\toprule
			Model & 1/25 & 1/26 & 1/27 & 1/28 & 1/29 & 1/30 & 1/31 & Avg. & \% Imp. \\ \midrule
			IRT & \textbf{0.1200} & \textbf{0.1090} & \textbf{0.1098} & \textbf{0.1184} & \textbf{0.1230} & \textbf{0.1311} & \textbf{0.1268} & \textbf{0.1199} \\
			LRT & 0.1375 & 0.1198 & 0.1203 & 0.1303 & 0.1347 & 0.1386 & 0.1347 & 0.1310 & 8.5\%\\
			Const & 0.1591 & 0.1361 & 0.1422 & 0.1510 & 0.1521 & 0.1631 & 0.1587 & 0.1520 & 21\%\\
			IR & 0.1396 & 0.1232 & 0.1291 & 0.1348 & 0.1396 & 0.1500 & 0.1425 & 0.1372 & 13\%\\
			LR & 0.1478 & 0.1262 & 0.1318 & 0.1418 & 0.1459 & 0.1567 & 0.1501 & 0.1431 & 16\%\\
			IRKM & 0.1307 & 0.1155 & 0.1182 & 0.1267 & 0.1318 & 0.1408 & 0.1346 & 0.1285 & 6.7\%\\
			LRKM & 0.1419 & 0.1208 & 0.1275 & 0.1371 & 0.1386 & 0.1498 & 0.1443 & 0.1373 & 13\%\\
			DSP & 0.1661 & 0.1662 & 0.1759 & 0.1605 & 0.1646 & 0.1724 & 0.1763 & 0.1689 & 29\%\\ \bottomrule
		\end{tabular}%
	\par
	\medskip
	{\footnotesize \textit{Note.} The column ``Avg.'' measures the average MSE across all seven days of the test set, and the column ``\% Imp.'' measures the percentage improvement (decrease) in average MSE from the IRT relative to each benchmark.}
	\end{subtable}
\end{table}

The IRT attains a lower MSE than all benchmarks for each of the 21 individual days of test data. The IRT achieves a 5-29\% improvement in overall MSE and 2-14\% improvement in AUC over the DSP's approach across the three exchanges. The IRT also achieves a 7-13\%/7-15\% improvement in MSE/AUC relative to the IR benchmark and a 1-7\%/0.6-5\% improvement relative to IRKM. The strong performance of IRT over IR demonstrates the value of segmentation in bid landscape forecasting. Moreover, the superior performance of IRT over IRKM illustrates the gains achieved by applying a supervised segmentation procedure, driven by accurately capturing differences in the underlying segments' bid landscapes. Notably, each benchmark using isotonic regression achieves better empirical performance than its logistic regression counterpart. This finding illustrates that isotonic regression models can offer substantial improvements in terms of predictive accuracy over other parametric approaches for bid landscape forecasting.

\begin{figure}[!htbp]
\centering
\caption{Exchange 1: In-sample and out-of-sample MSEs and AUCs of IRT and IRKM as a function of the number of auction segments.}\label{fig:perf-func-Kexch1}
\begin{subfigure}{.5\linewidth}
\centering
\includegraphics[width=0.99\linewidth]{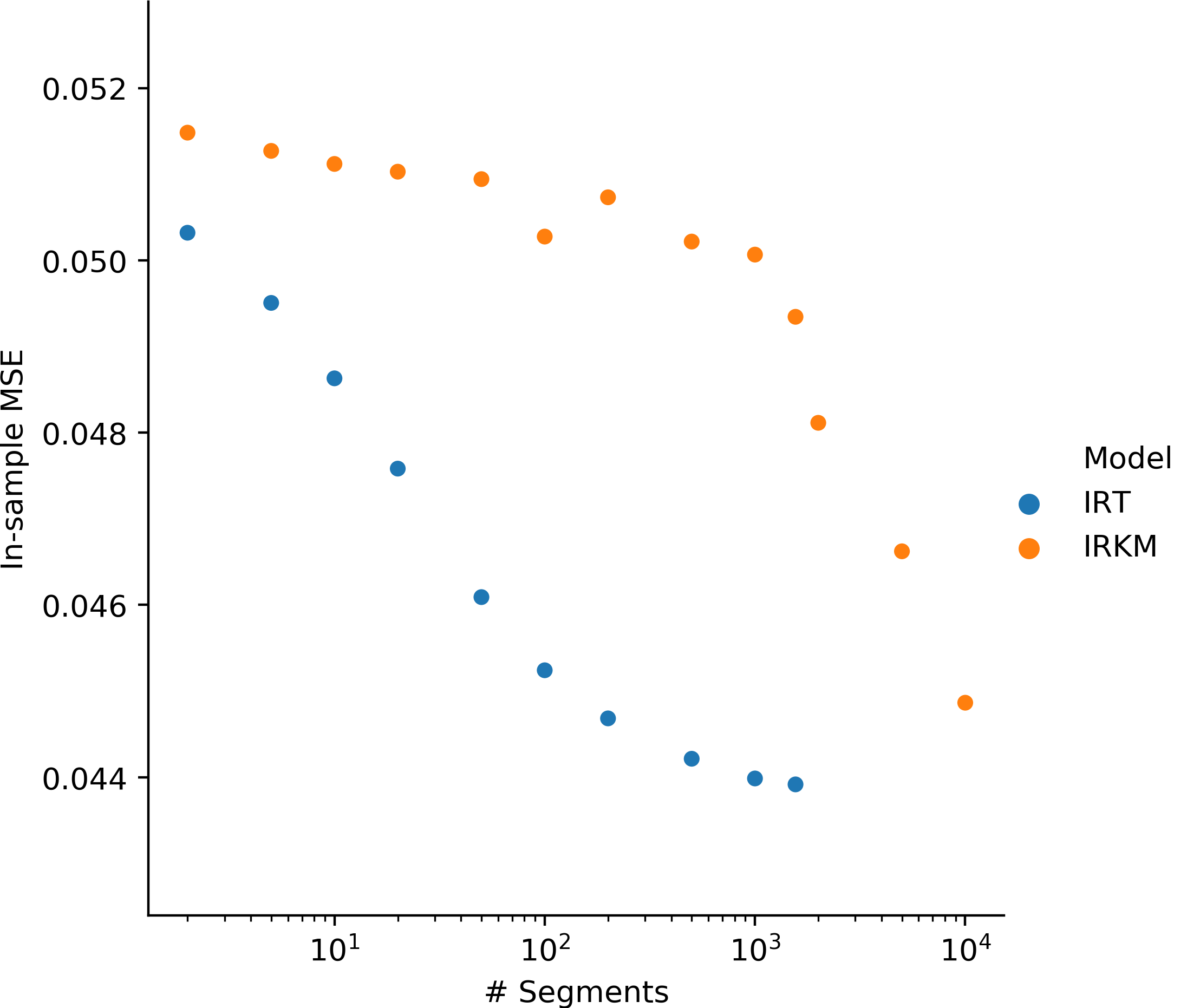}
\label{fig:mseexch1train}
\end{subfigure}%
\begin{subfigure}{.5\linewidth}
\centering
\includegraphics[width=0.99\linewidth]{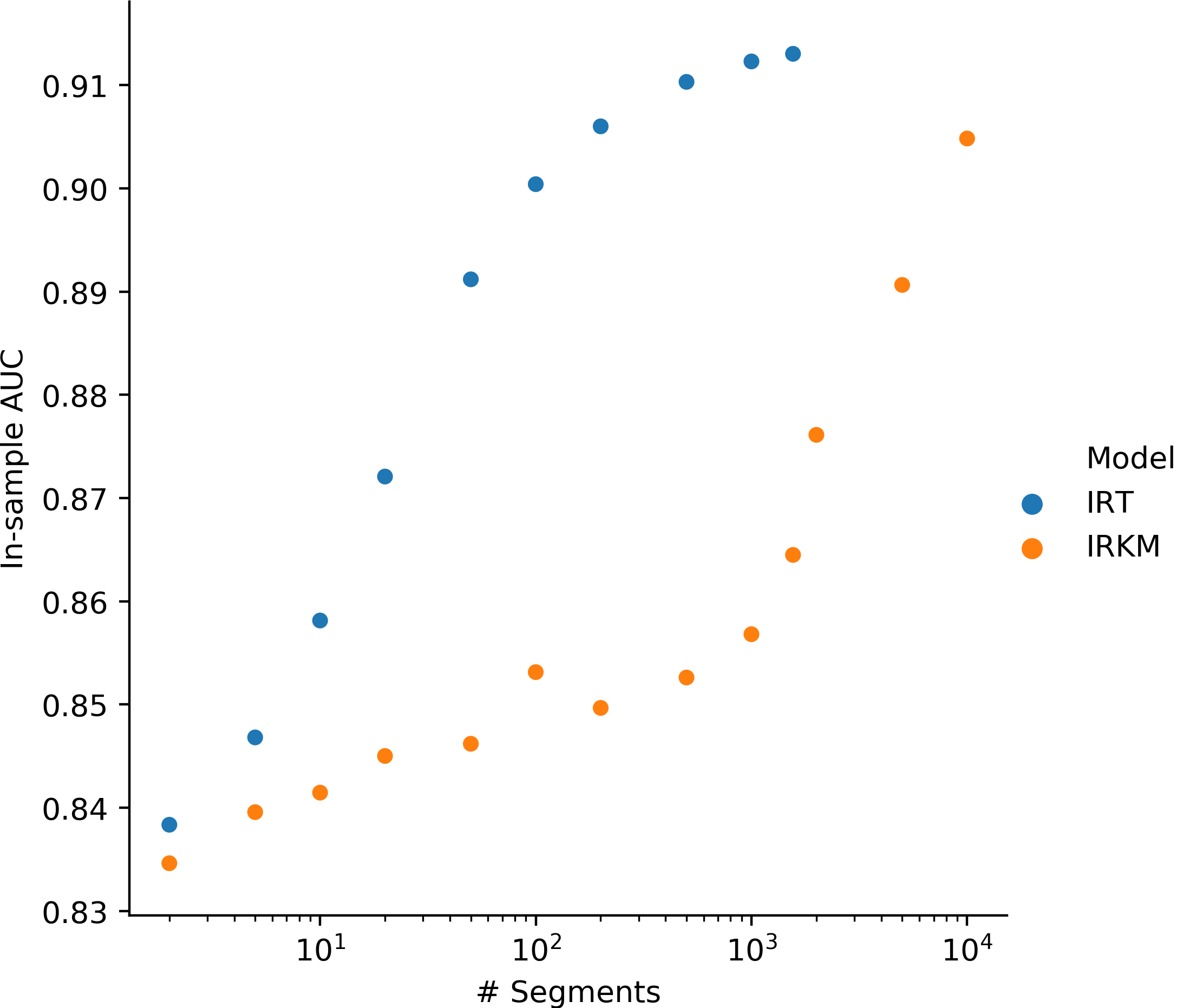}
\label{fig:aucexch1train}
\end{subfigure}

\begin{subfigure}{.49\linewidth}
\centering
\includegraphics[width=0.99\linewidth]{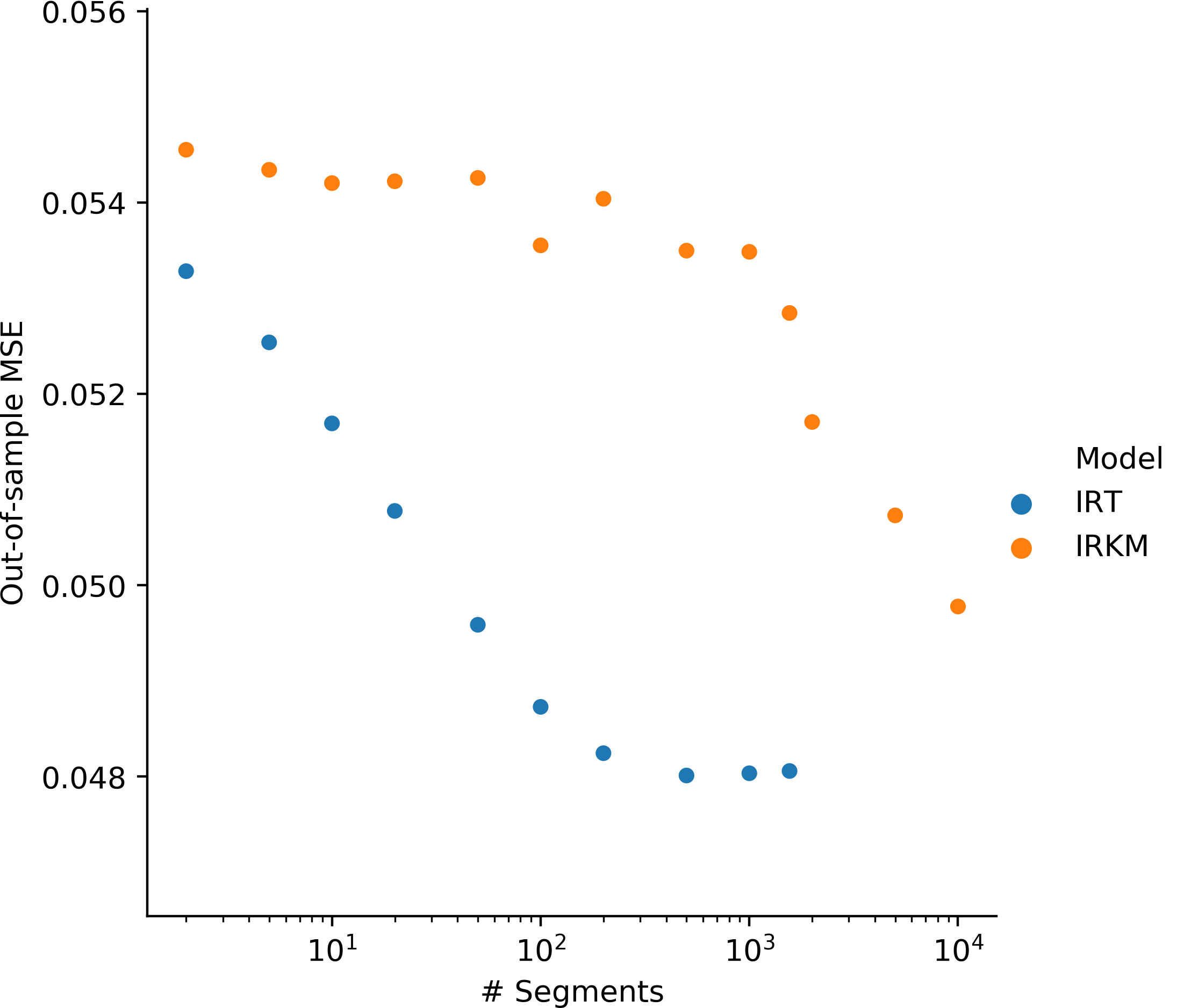}
\label{fig:mseexch1test}
\end{subfigure}
\begin{subfigure}{.49\linewidth}
\centering
\includegraphics[width=0.99\linewidth]{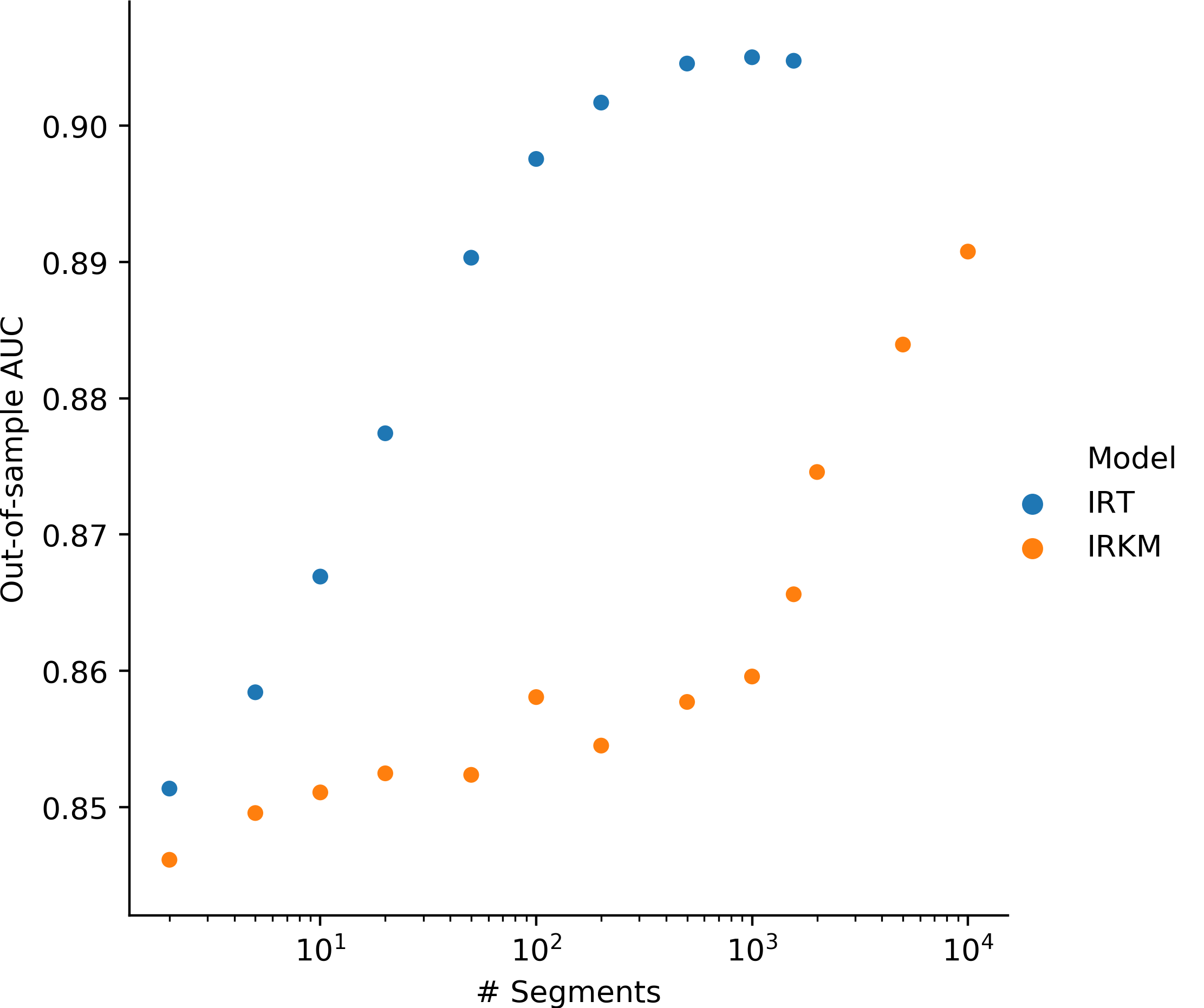}
\label{fig:aucexch1test}
\end{subfigure}
\end{figure}

\paragraph{Results: Interpretability.} 
The only algorithm on par with IRT for predictive performance is IRKM, so we next explore the interpretability of the IRT and IRKM algorithms.
Figure \ref{fig:perf-func-Kexch1} plots the MSE/AUC performance of the IRT and IRKM algorithms as a function of the number of segments for exchange 1. Similar plots for exchanges 2 and 3 are given in Appendix \ref{sec:DSPExp}. We observe across all 3 exchanges that IRTs using only 10 market segments perform better than IRKM models using up to 1000 market segments. These experiments demonstrate that IRTs are capable of constructing concise decision trees that yield equivalent predictive performance relative to IRKMs requiring the use of considerably more segments.
In addition, given the same number of market segments,  the IRT's tree-based market segmentation leads to considerably more interpretable market segments than $K$-means clustering, agreeing with the findings at the end of Section \ref{sec:Swissmetro}. {\color{black}Appendix \ref{sec:DSPInt} visualizes and interprets some of the key differences found in the IRT's bid landscape response models.}

\section{Conclusion}

We propose a new framework for tractably training decision trees for the purposes of market segmentation and personalized decision-making which we call Market Segmentation Trees (MSTs). While more traditional approaches to market segmentation (e.g., $K$-means) segment customers solely on the basis of their feature similarity, MSTs learn a market segmentation explicitly driven by identifying and grouping together customers with similar responses to decisions. We propose a training algorithm for MSTs in which decision tree splits are recursively selected to optimize the predictive accuracy of the resulting collection of response models. We provide an open-source code base in Python which implements the training algorithm and can be easily customized to fit different personalized decision-making applications. We incorporate several strategies into the code base for improved scalability such as parallel processing and warm starts, and we provide a theoretical analysis of the code's asymptotic computational complexity supporting its tractability in large data settings.

To demonstrate the versatility of our methodology, we design two new, specialized MST algorithms: $(i)$ Choice Model Trees (CMTs) which can be used to predict a user's choice amongst multiple options, and $(ii)$ Isotonic Regression Trees (IRTs) which can be used to solve the bid landscape forecasting problem. 
We examine the performance of CMTs on a variety of synthetic datasets, observing that CMTs reliably find market segmentations which accurately predict choice probabilities, overcome model misspecification, and are robust to overfitting.
\edit{We also apply our CMT algorithm to segment  users from the Swissmetro dataset, and we find that CMTs have strong predictive performance compared to common and state-of-the-art benchmarks, while providing an interpretable segmentation of users. }
We then examine the performance of IRTs using a large-scale dataset from a leading Demand Side Platform (DSP), where we segment advertisement opportunities for users in order to predict auction win rate as a function of bid price. Our IRT algorithm consistently outperforms all benchmarks across 21 individual days of test data. \edit{For both these datasets, the CMT and IRT methods only require a fraction of the number of market segments compared to the classic $K$-means clustering approach.}

\ACKNOWLEDGMENT{We greatly appreciate the feedback from the review team, all of whom  helped improve the paper significantly. AE was partially supported by NSF grant CMMI-1763000. 
}
\bibliographystyle{ormsv080}  
\bibliography{mst}

\newpage


\ECSwitch
\ECHead{Online Appendix: Market Segmentation Trees}

\begin{APPENDICES}

\counterwithin{thm}{section}
\counterwithin{lem}{section}
\counterwithin{asmpn}{section}
{\blockedit
\section{Proof of Theorem \ref{thm:bound1}} \label{sec:proofthm1}
 
The goal is to show that the computational complexity of training a depth $D$ tree under Assumptions \ref{asn:gprop1} and \ref{asn:Nprop} is $O\big(D\cdot m\cdot g(n)\big)$. By definition of Big O notation, this is equivalent to showing that there exits $N_0\geq 0$ and $C_0>0$ such that the computational complexity is at most $C_0 \cdot D\cdot m\cdot g(n)$ for any $n \geq N_0$, $m\geq 1$, and $D\geq 1$. Recall that $O(g(n))$ is an upper bound on the computational cost of fitting a response model to training data of size $n$, i.e., the cost of solving the optimization problem in Eq. (\ref{eqn:leafmodtrain}). Thus, there exists an $N_1 \geq 0$ and $C_1 > 0$ such that the cost of fitting the response model on $n$ observations is at most $C_1 \cdot g(n)$ for all $n \geq N_1$. From Assumption \ref{asn:Nprop}, we can and shall select $N_0$ such that $N_{\mathcal{T}}(D,l;n) \geq N_1$ for all $n\geq N_0$, $l=1,\ldots,2^D$, and all $\mathcal{T}$.

Our proof relies on the following result.

\begin{lemma}[\citet{bruckner1962some}] \label{lem:superadd}
\textit{Let $g(n)$ be a nonnegative, continuous, and convex function which satisfies $g(0) = 0$. Then,}
\begin{enumerate}
    \item[a)] \textit{$g(n)$ is star-shaped, i.e. $g(\alpha n) \leq \alpha g(n)$ for all $\alpha \in [0,1]$ and for all $n \geq 0$.}
    \item[b)] \textit{$g(n)$ is superadditive, i.e. $g(n_1+n_2) \geq g(n_1) + g(n_2)$ for all $n_1 \geq 0$ and $n_2 \geq 0$.}
\end{enumerate}
\end{lemma}
 
\noindent Thus, $g(n)$ is star-shaped and superadditive by Assumption \ref{asn:gprop1} and Lemma \ref{lem:superadd}.

To prove the theorem, we first analyze the computational complexity of the split selection procedure of Eq. (\ref{eqn:greedy}).  The lemma below bounds the runtime of the split selection procedure when applied in each internal node of the trained MST.

\begin{lemma} \label{lem:splitsel}
Suppose Assumptions \ref{asn:gprop1} and \ref{asn:Nprop} are satisfied and $n\geq N_0$. For all $\mathcal{T}$, $d \leq D-1$, $l \in \{1,...,2^d\}$,  the runtime of the split selection procedure with respect to $n$ observations and $m$ binary contextual variables is at most
\begin{equation*}  m C_1 g(N_{\mathcal{T}}(d,l;n)).
\end{equation*}
\end{lemma}

\proof{Proof.} To evaluate the quality of a candidate split, the split selection procedure fits response models within each of the two resulting partitions from the split and computes the cumulative training error across the partitions.  Let $n_1$ and $n_2$ denote the number of observations in each of the split's partitions, and note that $n_1 + n_2 = N_{\mathcal{T}}(d,l;n)$. By definition of $N_0$ above, $n_1 \geq N_1$ and $n_2 \geq N_1$. Thus, split evaluation takes time at most $C_1 g(n_1) + C_1 g(n_2)$ by the definition of $g(\cdot)$ and the fact that $n_1 \geq N_1$ and $n_2 \geq N_1$. It follows that
\begin{eqnarray*}
C_1 g(n_1) + C_1 g(n_2) &\leq C_1 g(N_{\mathcal{T}}(d,l;n))
\end{eqnarray*}
\noindent by the superadditivity of $g(\cdot)$. Since there are $m$ binary contextual variables, there are $m$ candidate splits which the split selection procedure must evaluate. Thus, the runtime for the split selection procedure is bounded by $mC_1 g(N_{\mathcal{T}}(d,l;n))$. \Halmos \endproof

The split selection procedure is recursively applied through all internal nodes of the trained MST $\mathcal{T}$. Thus, using Lemma \ref{lem:splitsel} the runtime of the training algorithm can be bounded as 
\begin{eqnarray*}
 mC_1 \sum_{d=0}^{D-1} \sum_{l=1}^{2^d} g(N_{\mathcal{T}}(d,l;n)) \leq mC_1 \sum_{d=0}^{D-1}g(n) = mC_1 D g(n).
\end{eqnarray*}

\noindent where the inequality follows from the superadditivity of $g(\cdot)$ (noting that $\sum_{l=1}^{2^d} N_{\mathcal{T}}(d,l;n) = n$). Setting $C_0=C_1$, we have now completed the proof of Theorem \ref{thm:bound1}. \Halmos

\section{Proof of Theorem \ref{thm:bound2}} \label{sec:proofthm2}
  
The goal is to show that the computational complexity of training a depth $D$ tree under Assumptions \ref{asn:gprop1} and \ref{asn:Nprop} is $O\big(\max\{D/Q,1\} \cdot m\cdot g(n)\big)$. By definition of Big O notation, this is equivalent to showing that there exits $N_0\geq 0$ and $C_0>0$ such that the computational complexity is at most $C_0 \cdot \max\{D/Q,1\}\cdot m\cdot g(n)$ for any $n \geq N_0$, $m\geq 1$, and $D\geq 1$. Recall that $O(g(n))$ is an upper bound on the computational cost of fitting a response model to training data of size $n$. Thus, there exists an $N_1 \geq 0$ and $C_1 > 0$ such that the cost of fitting the response model on $n$ observations is at most $C_1 \cdot g(n)$ for all $n \geq N_1$. Recall that $g(n)$ is star-shaped and superadditive by Assumption \ref{asn:gprop1} and Lemma \ref{lem:superadd}. From Assumption \ref{asn:Nprop}, we can and shall select $N_0$ such that $N_{\mathcal{T}}(D,l;n) \geq N_1$ for all $n\geq N_0$, $l=1,\ldots,2^D$, and all $\mathcal{T}$. From Assumption \ref{asn:equalprop}, there exists $N_2\geq 0$ and $C_2>0$ such that  $N_{\mathcal{T}}(D,l;n) \leq C_2 n/2^d$ for all $n\geq N_2$. From Assumption \ref{asn:gprop2}, there exists $N_3\geq 0$ and $C_3 >0$ such that $g(C_2 n) \leq C_3 g(n)$ for all $n\geq N_3$. If $N_0 < \max\{N_2,N_3\}$, then increase $N_0$ to $\max\{N_2,N_3\}$ and otherwise keep it as is.

We now bound the runtime of the split selection procedure of Eq. (\ref{eqn:greedy}) when there are $n$ observations and $m$ binary contextual variables. Lemma \ref{lem:splitselB} below bounds the split selection procedure's runtime for each internal node of the trained MST $\mathcal{T}$.

\begin{lemma} \label{lem:splitselB}
Suppose Assumptions   \ref{asn:gprop1}, \ref{asn:Nprop},   \ref{asn:equalprop}, and \ref{asn:gprop2} hold and $n\geq N_0$. Then for all $d \leq D-1$ and $l \in \{1,...,2^d\}$,  the runtime of the split selection procedure of Eq. (\ref{eqn:greedy}) when there are $n$ observations and $m$ binary contextual variables is at most
\begin{equation*}
m C_1 C_3 g(n) /2^d  \,\,.
\end{equation*}
\end{lemma}

\proof{Proof.} We apply Lemma \ref{lem:splitsel} with Assumptions  \ref{asn:gprop1} and \ref{asn:Nprop} to see that the runtime of the split selection procedure when there are $n$ observations and $m$ binary contextual variables is at most
\begin{eqnarray*}
 m C_1 g(N_{\mathcal{T}}(d,l;n)) &\leq&  m C_1 g( C_2 n/2^d) \\
&\leq& m C_1 g(C_2 n)/2^d  \\
&\leq& m C_1 C_3 g(n) /2^d.
\end{eqnarray*}

\noindent Above, the first inequality follows from the definition of $C_2$ (Assumption \ref{asn:equalprop}) and the fact that $f$ increasing from Assumption \ref{asn:gprop1}.  The second inequality follows from the fact that $g(\cdot)$ is star-shaped. The third inequality follows from the definition of $C_3$ (Assumption \ref{asn:gprop2}). \Halmos \endproof

The split selection procedure is applied to each internal node  of the MST for $d \in \{0,...,D-1\}$ and for $l \in \{1,...,2^d\}$. We next bound the runtime of applying the split selection procedure to all nodes $l$ at a given depth $d$. Recall that our training algorithm parallelizes these $2^d$ procedures across the available computational cores $Q$. The total runtime of this parallelization scheme is upper bounded by the following job scheduling process. Assume that the $2^d$ split selection procedures (``jobs'') are run in batches of $Q$ (one job per core), and the next batch of $Q$ jobs are run only when all jobs in the current batch have terminated. There would then be $\Bigl\lceil \dfrac{2^d}{Q} \Bigr\rceil$ total batches, and the runtime of each individual job (and thus each batch) can be bound by Lemma \ref{lem:splitselB}. Thus, the runtime of parallelizing all $2^d$ split selection procedures at depth $d$ can be bounded by
\begin{equation*}
    \Bigl\lceil \dfrac{2^d}{Q} \Bigr\rceil m C_1 C_3 g(n) /2^d .
\end{equation*}

\noindent  Finally, the runtime of the MST's training procedure is equal to the runtimes of the split selection procedures across all depths $d \in \{0,...,D-1\}$ of the MST, which can be bounded as follows:
\begin{align*}
 \sum_{d=0}^{D-1}    \Bigl\lceil \dfrac{2^d}{Q} \Bigr\rceil m C_1 C_3 g(n) /2^d   &\leq \sum_{d=0}^{D-1} \left( \dfrac{2^d}{Q} +1 \right) m C_1 C_3 g(n) /2^d  \\
 &= \sum_{d=0}^{D-1} \frac{ m C_1 C_3 g(n) }{Q} + \frac{m C_1 C_3 g(n)}{2^d}  \\
 &\leq \frac{ D m C_1 C_3 g(n) }{Q}+ 2  m C_1 C_3 g(n)   \\
 &\leq 3 \max\{D/Q,1\} m C_1 C_3 g(n).
\end{align*}
Letting $C_0=3 C_1 C_3$ completes the proof of Theorem \ref{thm:bound2}. \Halmos 

}

\section{Details of Datasets Used in Section \ref{sec:experimental}} \label{sec:details}

\subsection{Details of Dataset Generation for Section \ref{sec:synthetic}}
\label{app:details-synthetic}

Below we provide details on how each dataset is generated for each of the three ground truth models summarized in Section \ref{sec:synthetic}.

\textit{Context-Free MNL:}
We generate the MNL's parameter vector $\beta$ by sampling each element of $\beta$ independently from a Uniform(-1,1) distribution. This MNL model is used to generate the choices for all users in the dataset. Each user is encoded by four contextual variables sampled independently from a Uniform(0,1) distribution. The number of options offered to each user is sampled uniformly-at-random from the set $\{2,3,4,5\}$, and each option is encoded by four features which are sampled independently from a Uniform(0,1) distribution for each user. Choices are simulated from the probability distribution specified by the MNL model given the assortment -- in particular, the contextual variables are \textit{not} considered when generating choices.


\textit{Choice Model Tree:}
First, the number of leaf nodes is sampled uniformly-at-random from the set $\{4,5,6,7\}$. Then, a CMT of depth at most three is randomly constructed which has the target number of leaf nodes. Recall that each (numeric) split of a CMT is encoded by a splitting variable and split point (e.g., ``$x_3 < 0.4$''). All splitting variables and split points contained in the CMT are sampled uniformly-at-random with the constraint that each split is roughly “balanced”, defined as the left and right children of the split containing at least 30\% of the contexts mapped to their parent. Each leaf contains an MNL instance whose parameter vector $\beta$ is generated by sampling each element of $\beta$ independently from a Uniform(-1,1) distribution. 
Contexts and options are generated in the same manner as they were for the Context-Free MNL ground truth model, with contextual features and options being sampled independently from a Uniform(0,1) distribution. Choices are generated for each user by (1) mapping the user to the leaf of the CMT corresponding to the user's context, and (2) sampling a choice from the user's offered assortment using the leaf's MNL model.

\textit{$K$-Means Clustering Model:}
First, the number of clusters $K$ is sampled uniformly-at-random from the set of values $\{4,5,6,7\}$; recall that we also used this set of values to sample the number of leaves present in the CMT ground truth model. Each cluster $k \in \{1,...,K\}$ has an associated MNL model whose parameter vector $\beta_k$ is generated by sampling each element of $\beta_k$ independently from a Uniform(-1,1) distribution. Furthermore, each cluster also has an associated “mean context vector” $\bar{x}_k$ whose entries are sampled independently from a Uniform(0,1) distribution. We next define a probability mass function (p.m.f.) $\pi = \{\pi_1,...,\pi_K\}$ over the $K$ clusters, where $\pi_k$ denotes the probability that a user belongs to cluster $k$. We generate the p.m.f. through the following procedure:

\begin{enumerate}
    \item For each cluster $k \in \{1,...,K\}$, sample $U_k \in \mathbb{R}$ from a Uniform(-1,1) distribution.
    \item Let $\pi_k := \dfrac{\exp(U_k)}{\sum_{k'=1}^K \exp(U_{k'})}$ for all $k \in \{1,...,K\}$.
\end{enumerate}

Options are generated through the same procedure as in the other two ground truth models, with option features being sampled independently from a Uniform(0,1) distribution. Contexts and choices are generated for each user in the following manner:

\begin{enumerate}
    \item Sample the cluster $k \in \{1,...,K\}$ belonging to the user from p.m.f. $\pi$.
    \item Sample the user's context vector from a multivariate normal distribution with mean parameter $\bar{x}_k$ and covariance $\sigma^2 I$, where $I$ denotes the identity matrix. Here, $\sigma = 0.08$ is configured to ensure that there is an adequate separation between contexts belonging to different clusters.
    \item Sample the user's choice from the MNL model associated with cluster $k$, i.e. the MNL model with parameter vector $\beta_k$.
\end{enumerate}

{\color{black}
\subsection{Description of Swissmetro Dataset for Section \ref{sec:Swissmetro}} \label{app:details-swiss}
A detailed description of the dataset as well as the data collection procedure is given in~\citet{bierlaire2001acceptance}.

There are 11 features used as contexts for segmentation which can be categorized as follows. Variable names as listed in~\citet{bierlaire2001acceptance} are included in caps in parentheses. All features except age are treated as categorical (not ordinal), whereas age is ordinal.
\begin{itemize}
\item \emph{Demographic information:} gender (MALE), age (AGE), and annual income (INCOME). We used the data exactly as provided, with the exception that we combined INCOME=0 and INCOME=1 because they both referred to the same annual income.

\item \emph{Information regarding use of transportation:} travel purpose such as traveling for business or shopping (PURPOSE), whether the person is a current rail user (GROUP), whether the person has an annual rail pass (GA), whether the person prefers first class travel (FIRST), ticket type (TICKET), party responsible for paying for ticket such as the person or employer (WHO), and how much luggage the person is bringing (LUGGAGE). We used the data exactly as provided.

\item \emph{Route information:} the data included the origin (ORIGIN) and destination (DEST) canton of each route. Using this data, we calculated the number of observations with routes between the same two cantons; if the number of observations was at least 250, we included a binary feature consisting of all routes between the two cantons (11 binary features in total).
\end{itemize}

As described in Section \ref{sec:Swissmetro}, people who owned a car had three options of transportation modes to choose from - car, train, and Swissmetro - whereas people who did not own a car chose between two options - train and Swissmetro. We removed observations from the dataset with no reported choices. Each option is described by the following three features and an alternative-specific fixed effect. Variable names as listed in~\citet{bierlaire2001acceptance} are included in caps in parentheses. 

\begin{itemize}
\item Door-to-door travel times in minutes (TRAIN\_TT, SM\_TT, CAR\_TT).
\item Cost in CHF (TRAIN\_CO, SM\_CO, CAR\_CO). The cost of the car option was estimated considering a fixed average cost per kilometer.
\item Headway in minutes, i.e., the average time between two consecutive train/Swissmetro arrivals (TRAIN\_HE, SM\_HE). We set the car headway equal to zero.
\end{itemize}


}

\subsection{Description of DSP Dataset for Section \ref{sec:DSP}}
\label{app:details-DSP}

There are ten user and ad spot auction features used as contexts for segmentation which can be categorized as follows:
\begin{itemize}
    \item \textit{Information regarding the ad spot}: Area and aspect ratio of the ad spot (defined as ``width$\times$height'' and ``width/height'', respectively), ad spot fold position (defined as whether the ad is visible without scrolling), and ID of the encompassing site. Area and aspect ratio are treated as numeric features in the IRT; all other reported features are treated as categorical. Due to the high dimensionality of the site IDs (with thousands of unique values per exchange), we first pre-cluster the site IDs before applying the IRT and the benchmark algorithms to the training data.
    \item \textit{Information regarding the user's site visit}: Time-of-day and day-of-week of the user's site visit, country of the visiting user, and ad channel from which the user arrived (e.g., video, mobile, search).
    \item \textit{Information regarding private marketplace deals}: ID encoding a private deal between an advertiser and a publisher which might affect the dynamics of the auction.
\end{itemize}

{\color{black}

\section{Additional Results for CMT Performance Evaluation from Section~\ref{sec:Swissmetro}} \label{app:AddCMT}

We implement and evaluate two benchmarks that do \emph{not} provide a market segmentation in the form of a finite partition of the population. Instead, the response models use both contextual features ($x$) and decision variables ($p_h$) to predict users' responses.
\begin{itemize}
    \item {\em MNL-INT:} We augment the context-free MNL model by accounting for all pairwise interactions between contextual variables and decision variables. 
Letting $I_{i,h}$ be the vector of pairwise interactions between binarized $x_i$ and $p_{i,h}$ for all $i \in [n]$, MNL-INT uses the utility function $U_{i,h} = \beta^T p_{i,h} + \gamma^T  I_{i,h} +  \epsilon_{i,h}$, where $\beta$ and $\gamma$ are the parameter vectors. 


\item {\em Taste-Net:} We implement the neural network architecture for choice modelling developed by a subsequent paper~\citet{han2020neural}, which is called Taste-Net. This approach specifies the random utility  $U_{i,h} = g(x_i)^T p_{i,h} +   \epsilon_{i,h}$ for all $i \in [n]$, where the coefficient $g(x_i)$ is the output of a deep feed-forward neural network. The training objective is to minimize the negative log-likelihood loss of the travel mode choices. 
\end{itemize}

MNL-INT and Taste-Net are implemented using the Adam gradient-descent optimizer of the Tensorflow library.
For both benchmarks, we pick the learning rate and terminal epoch (known as ``early stopping'') of the gradient-descent in $\{0.001,0.0001\}\times [1,1000]$ by tracking the log-likelihood loss on the validation set.

To enable fair comparisons, we also implement modified versions CMT+, MNL-KM+, MNL-DT+, and MNL-ICOT+, where the MNL response models incorporate contextual information. Specifically, each context feature is weighted by an alternative-specific coefficient in the utility function. This approach yields greater expressive power, at the cost of a less interpretable market segmentation. Indeed, the probabilistic response is  personalized for each user context,  beyond its segment membership.

Table~\ref{tab:swissresults2} shows the test set performance of these methods for each of the 10 random splits. Similarly to our previous findings, CMT+ achieves higher predictive accuracy on average than other methods. 
Taste-Net concedes an average $6.5\%$ NLL gap and $2.7\%$ MSE gap against CMT+.  
 Finally, from Table \ref{tab:expresults} and Table \ref{tab:swissresults2} we note that CMT+ outperforms CMT by 9.6\% in NLL and 9.2\% in MSE.

\begin{table}[H]
\caption{Test set negative log-likelihoods (NLL) and mean squared errors (MSE) of the tested methods on 10 random splits of the data, labeled as S1 through S10.}
\label{tab:swissresults2}
	\begin{subtable}{1\linewidth}
	\vspace{0.2cm}
	\caption{Test set average NLL}
	\vspace{0.05cm}
	\resizebox{\textwidth}{!}{%
	    \setlength\tabcolsep{2pt}
		\begin{tabular}{@{}llllllllllllr@{}}
			\toprule
			Model & S1 & S2 & S3 & S4 & S5 & S6 & S7 & S8 & S9 & S10 & Avg. & \% Imp. \\ \midrule
CMT+&0.5504&{ 0.5443}&{\bf 0.5536}&{\bf 0.5910}&{\bf 0.5546}&{\bf 0.5986}&{ 0.5598}&{\bf 0.5437}&{\bf 0.5437}&0.5548&{\bf 0.5595}& -\\
MNL-INT&0.6977&0.6958&0.7018&0.6993&0.6866&0.7061&0.7271&0.6856&0.6884&0.6933&0.6982&19.9\%\\
MNL-KM+&{\bf 0.5483}&{\bf 0.5204}&0.6403&0.5999&0.5654&0.6355&{\bf 0.5473}&0.5892&0.5796&{\bf 0.5065}&0.5732&2.4\%\\
MNL-DT+&0.6137&0.5697&0.5922&0.6027&0.6056&0.6190&0.5807&0.5637&0.5628&0.5384&0.5848&4.3\%\\
MNL-ICOT+&0.5759&0.5466&0.5861&0.6005&0.5865&0.6048&0.5866&0.5784&0.5660&0.5499&0.5781&3.2\%\\
Taste-Net&0.5936&0.5639&0.6299&0.6407&0.6063&0.6139&0.5759&0.5897&0.6008&0.5700&0.5985&6.5\%\\
			 \bottomrule
		\end{tabular}
	}
	\end{subtable}

	\begin{subtable}{1\linewidth}
	\vspace{0.2cm}
	\caption{Test set MSE}
	\vspace{0.05cm}
	\resizebox{\textwidth}{!}{%
	    \setlength\tabcolsep{2pt}
		\begin{tabular}{@{}llllllllllllr@{}}
			\toprule
			Model & S1 & S2 & S3 & S4 & S5 & S6 & S7 & S8 & S9 & S10 & Avg. & \% Imp. \\ \midrule
CMT+&{\bf 0.3204}&{\bf 0.3283}&{\bf 0.3342}&{\bf 0.3467}&{\bf 0.3306}&{\bf 0.3475}&{\bf 0.3259}&{\bf 0.3263}&{\bf 0.3191}&0.3301&{\bf 0.3309}&-\\
MNL-INT&0.4219&0.4183&0.4261&0.4252&0.4141&0.4300&0.4360&0.4117&0.4140&0.4161&0.4213&21.5\%\\
MNL-KM+&0.3503&0.3362&0.3580&0.3631&0.3558&0.3608&0.3572&0.3516&0.3400&0.3331&0.3506&5.6\%\\
MNL-DT+&0.3460&0.3485&0.3577&0.3575&0.3595&0.3569&0.3430&0.3357&0.3361&0.3246&0.3466&4.5\%\\
MNL-ICOT+&0.3503&0.3362&0.3580&0.3631&0.3558&0.3608&0.3572&0.3516&0.3400&0.3331&0.3506&5.6\%\\
Taste-Net&0.3312&0.3289&0.3502&0.3581&0.3549&0.3519&0.3320&0.3347&0.3348&{\bf 0.3248}&0.3402&2.7\%\\
			\bottomrule
		\end{tabular}%
	} \par
	\medskip
	{\footnotesize \textit{Note.} The column ``Avg.'' measures the average error across all 10 splits, and the column ``\% Imp.'' measures the percent improvement of CMT's average error compared to that of the benchmark.} 
	\end{subtable}
\end{table}

\clearpage \section{Example Tree Produced by CMT in Section \ref{sec:Swissmetro}} \label{app:CMTS10}

Figure \ref{fig:CMTS10} displays the tree defining market segments for split S10, as an example of a tree produced by the CMT method with tree depth 4. Each node contains a binary question, together which partition the set of possible observations into 16 market segments. Each market segment can be fully described by at most four features. For example, market segment 1 consists of current rail users traveling for business who are less than 25 years old.

\begin{figure}[h]
    \centering
	\includegraphics[width=16cm]{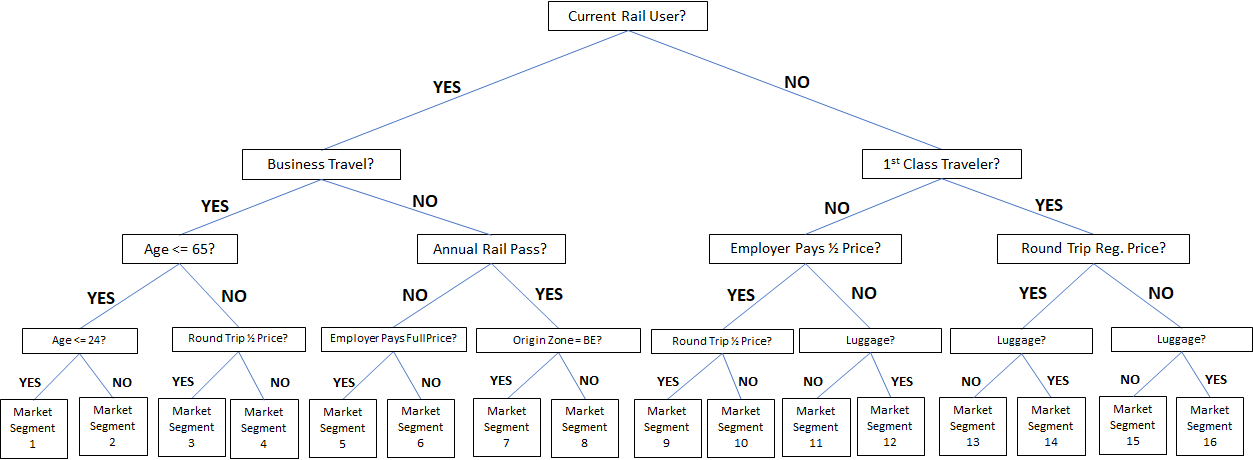}
	\caption{Tree defining market segments produced by the CMT with depth 4 for split S10.}
	\label{fig:CMTS10}
\end{figure}

{\color{black}To gain intuition as to how utility model parameters differ across market segments, one can compare the values of option-specific parameters $\beta_{l,h}$ across different market segments $l$ and for each option $h$. To illustrate, Figure \ref{fig:betasS10} provides a scatter plot showing each market segment's betas for Swissmetro's cost and travel time. As expected, nearly all of these betas are negative, implying that people prefer the Swissmetro option to have a shorter travel time and lower cost\footnote{Market segment 10 is an exception but it could be an effect of collinearity between travel time and cost.}.
As an example of comparing betas across market segments, consider segments 1 and 2, which together represent the subset of people who are current rail users and travel for business; for this subset of people, we find that young people (market segment 1: age $\leq$ 24) are relatively more sensitive to cost and less sensitive to travel time than middle-aged people (market segment 2: 25 $\leq$ age $\leq$ 65).}

\begin{figure}[h]
    \centering
	\includegraphics[width=10cm]{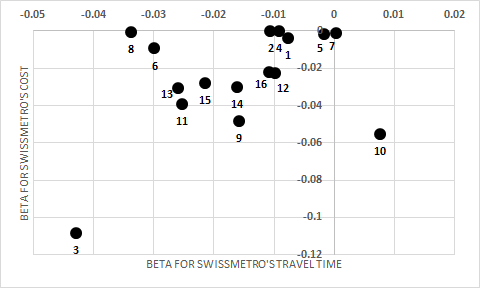}
	\caption{{\color{black}Scatter plot of betas for Swissmetro's cost and travel time, with each dot representing a pair of betas for its labeled market segment produced by the CMT with depth 4 for split S10.}}
	\label{fig:betasS10}
\end{figure}

\clearpage \section{Additional Figures for IRT Performance Evaluation from Section \ref{sec:DSP}} \label{sec:DSPExp}

In Section \ref{sec:DSP}, we plot the bid landscape forecasting performance of the IRT and IRKM algorithms as a function of the number of auction segments for ad exchange 1. Figures \ref{fig:perf-func-Kexch2} and \ref{fig:perf-func-Kexch3} provide plots for ad exchanges 2 and 3, which show similar insights.

\begin{figure}[!htbp]
\centering
\caption{Exchange 2: In-sample and out-of-sample MSEs and AUCs of IRT and IRKM as a function of the number of auction segments.}\label{fig:perf-func-Kexch2}
\begin{subfigure}{.5\linewidth}
\centering
\includegraphics[width=0.99\linewidth]{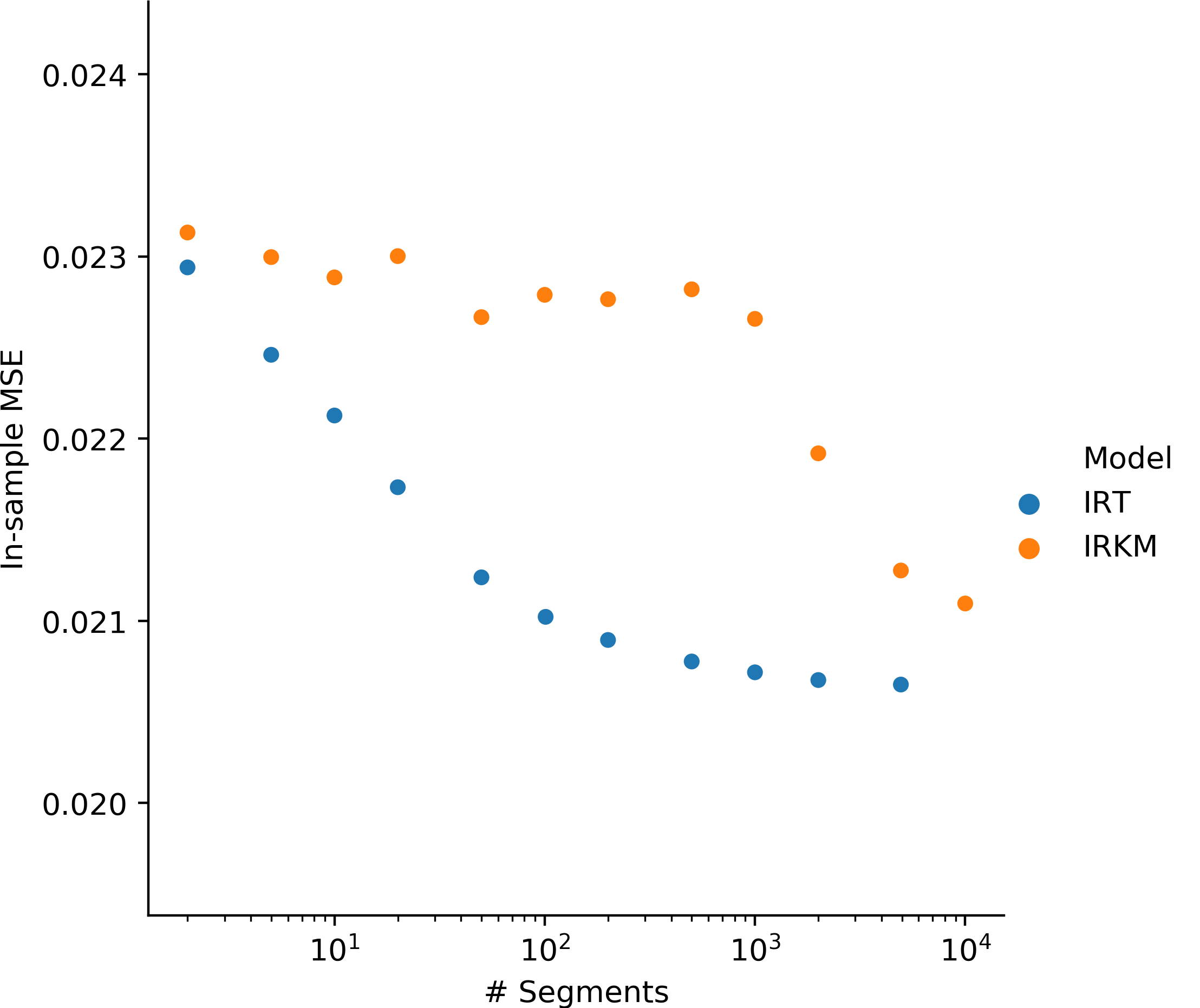}
\end{subfigure}%
\begin{subfigure}{.5\linewidth}
\centering
\includegraphics[width=0.99\linewidth]{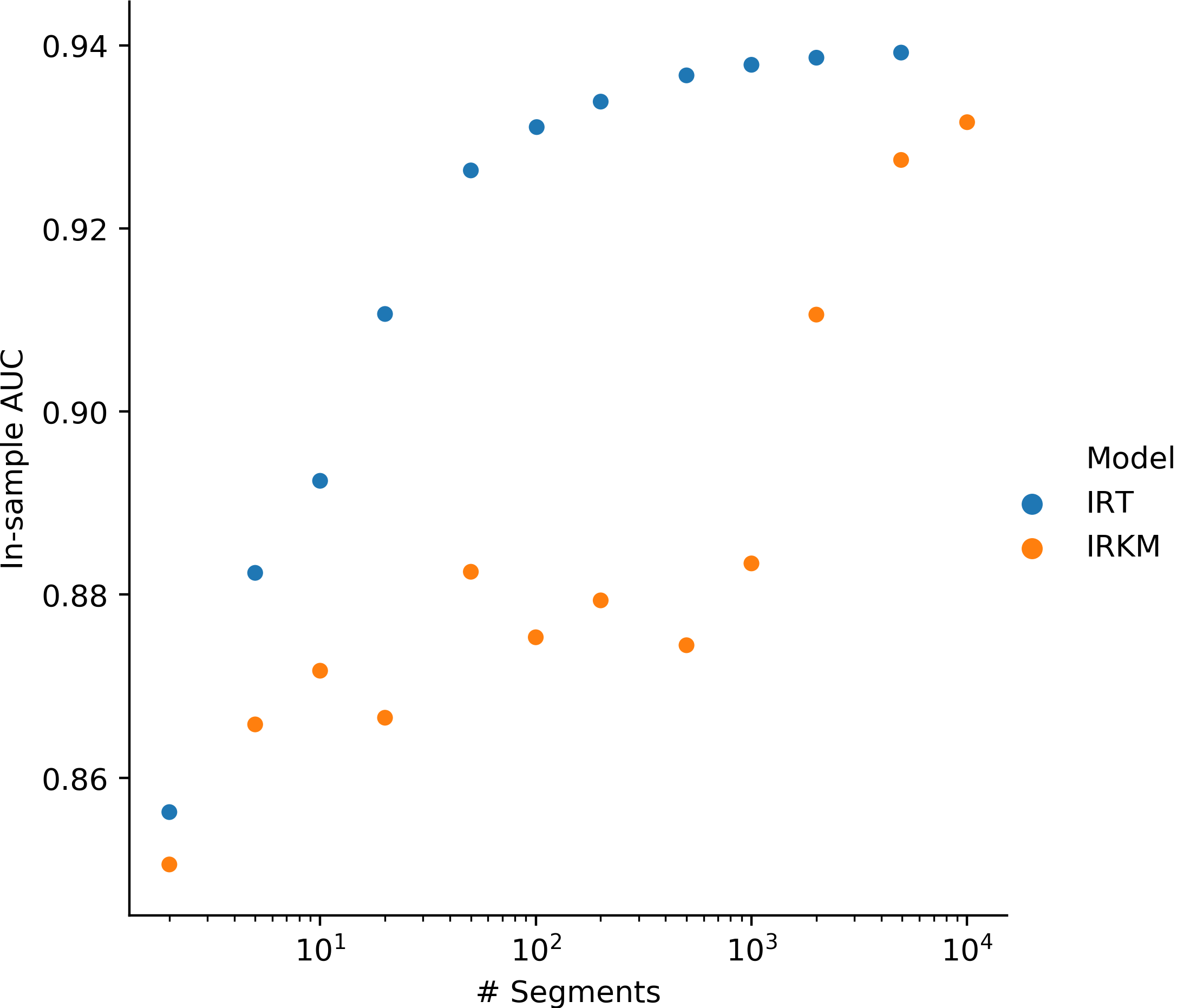}
\end{subfigure}

\begin{subfigure}{.49\linewidth}
\centering
\includegraphics[width=0.99\linewidth]{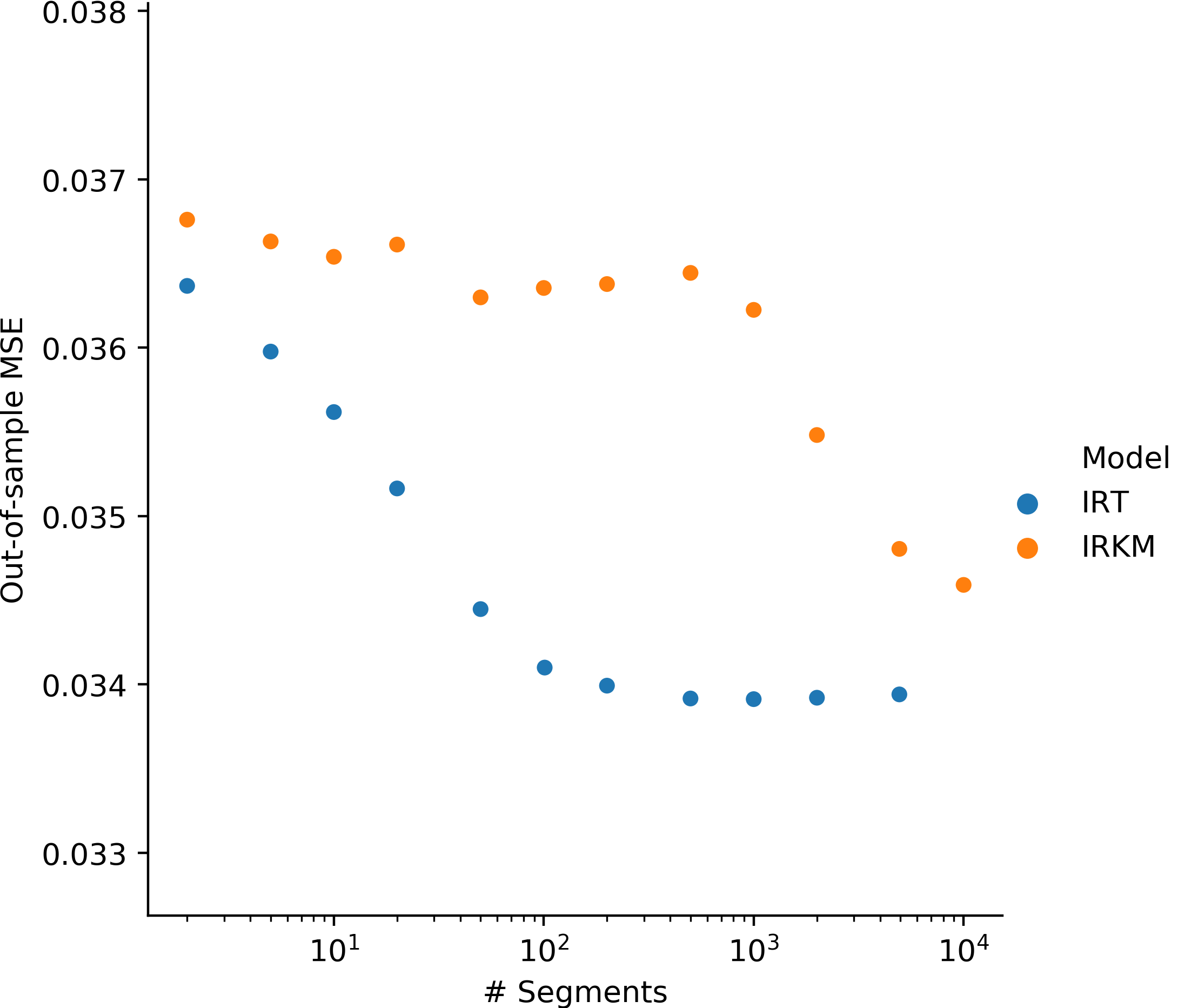}
\label{fig:mseexch2test}
\end{subfigure}
\begin{subfigure}{.49\linewidth}
\centering
\includegraphics[width=0.99\linewidth]{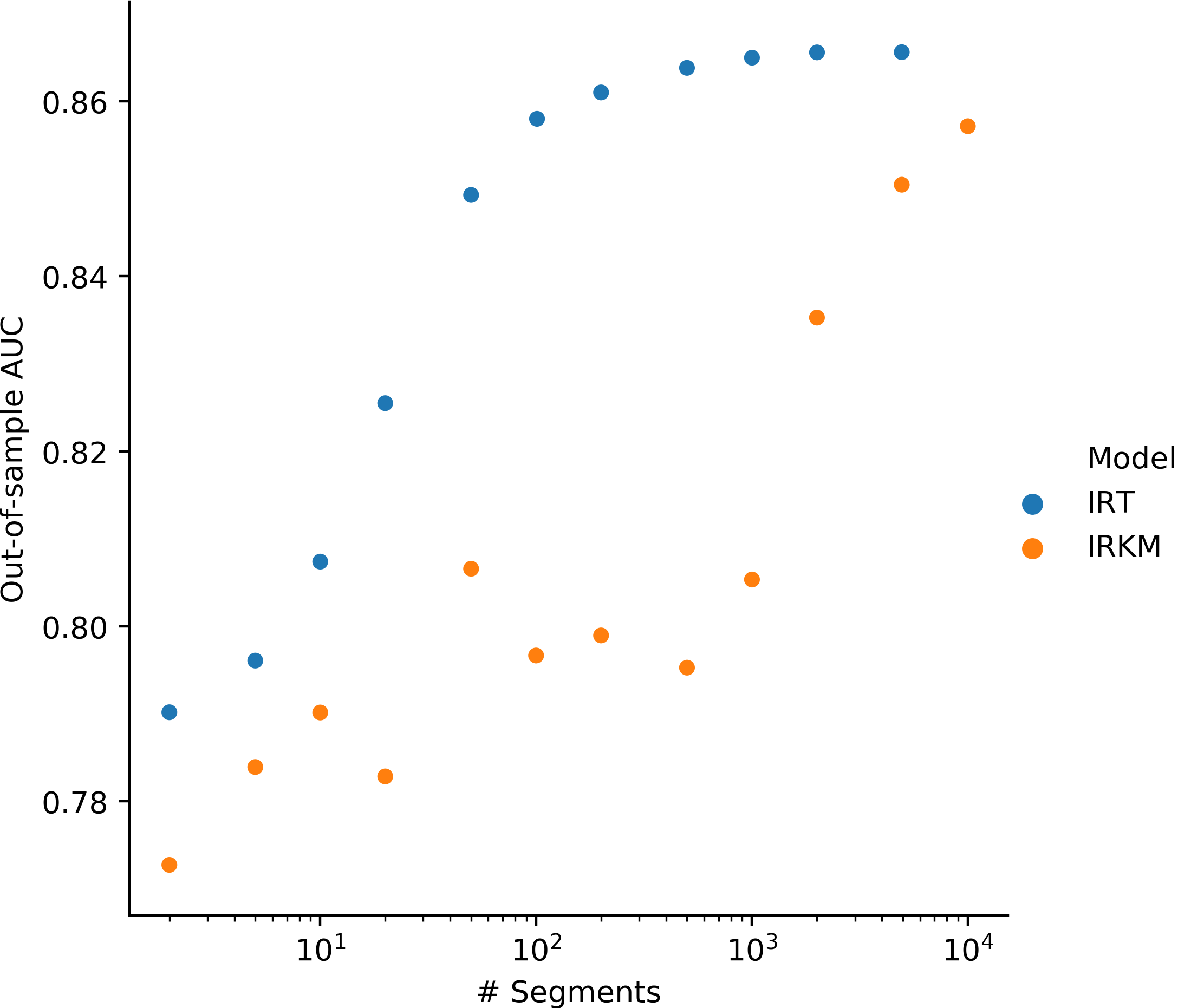}
\label{fig:aucexch2test}
\end{subfigure}
\par
	\medskip
	{\small  \textit{Note.} For IRT, we vary the number of segments from 1 to 4942. For IRKM, we vary the number of segments from 1 to 10000.}
\end{figure}

\begin{figure}[!htbp]
\centering
\caption{Exchange 3: In-sample and out-of-sample MSEs and AUCs of IRT and IRKM as a function of the number of auction segments.}\label{fig:perf-func-Kexch3}
\begin{subfigure}{.5\linewidth}
\centering
\includegraphics[width=0.99\linewidth]{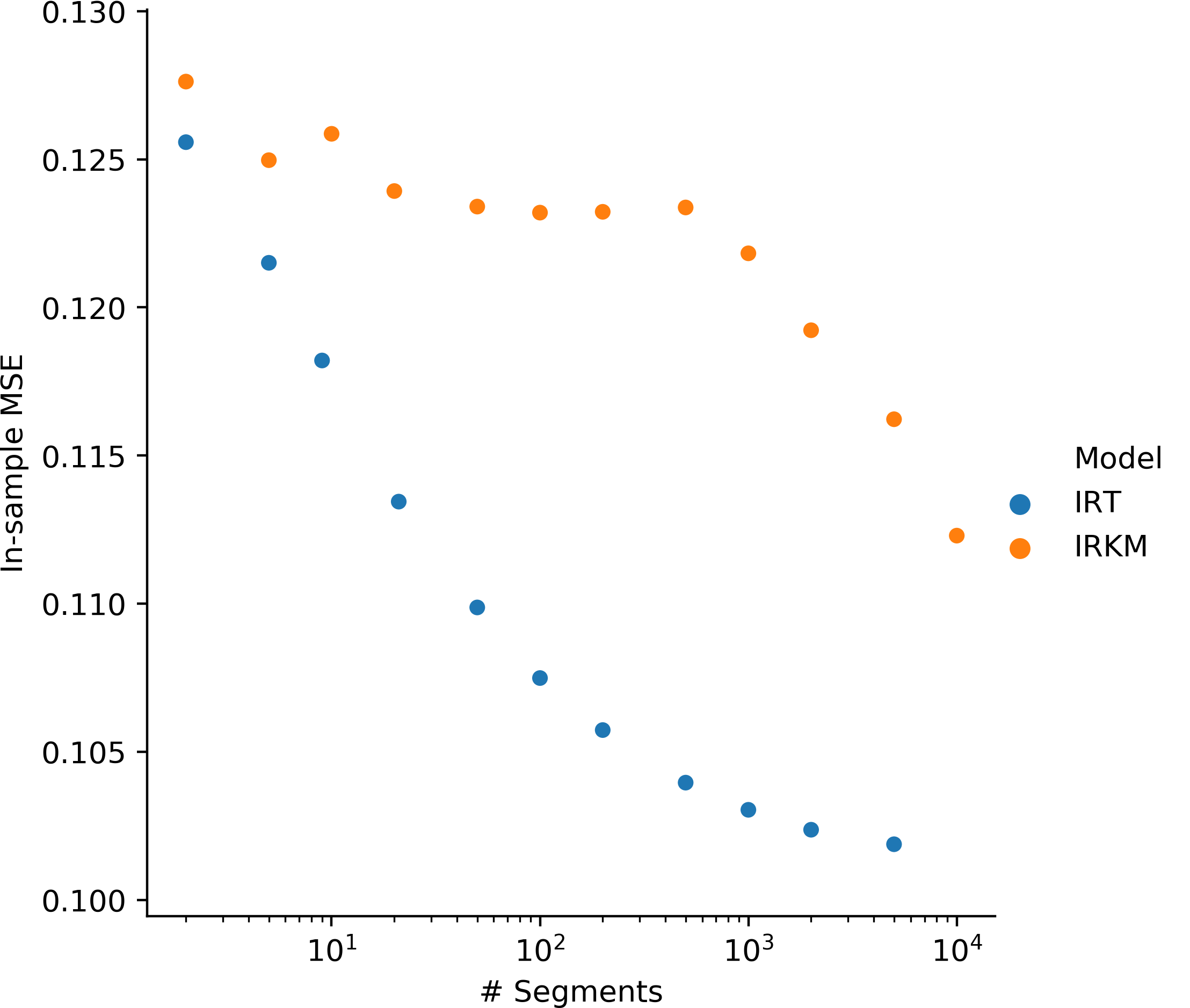}
\label{fig:mseexch3train}
\end{subfigure}%
\begin{subfigure}{.5\linewidth}
\centering
\includegraphics[width=0.99\linewidth]{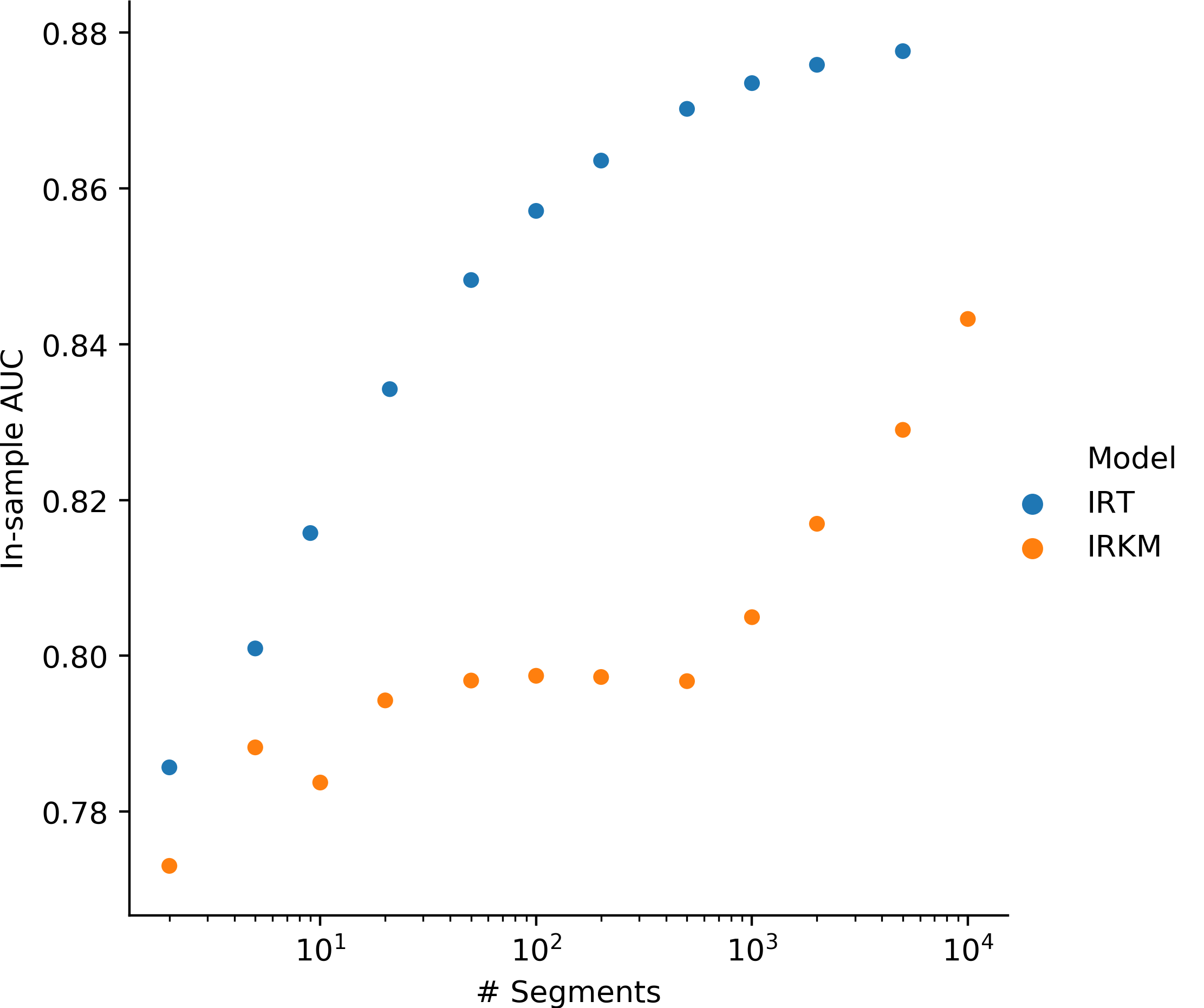}
\label{fig:aucexch3train}
\end{subfigure}

\begin{subfigure}{.49\linewidth}
\centering
\includegraphics[width=0.99\linewidth]{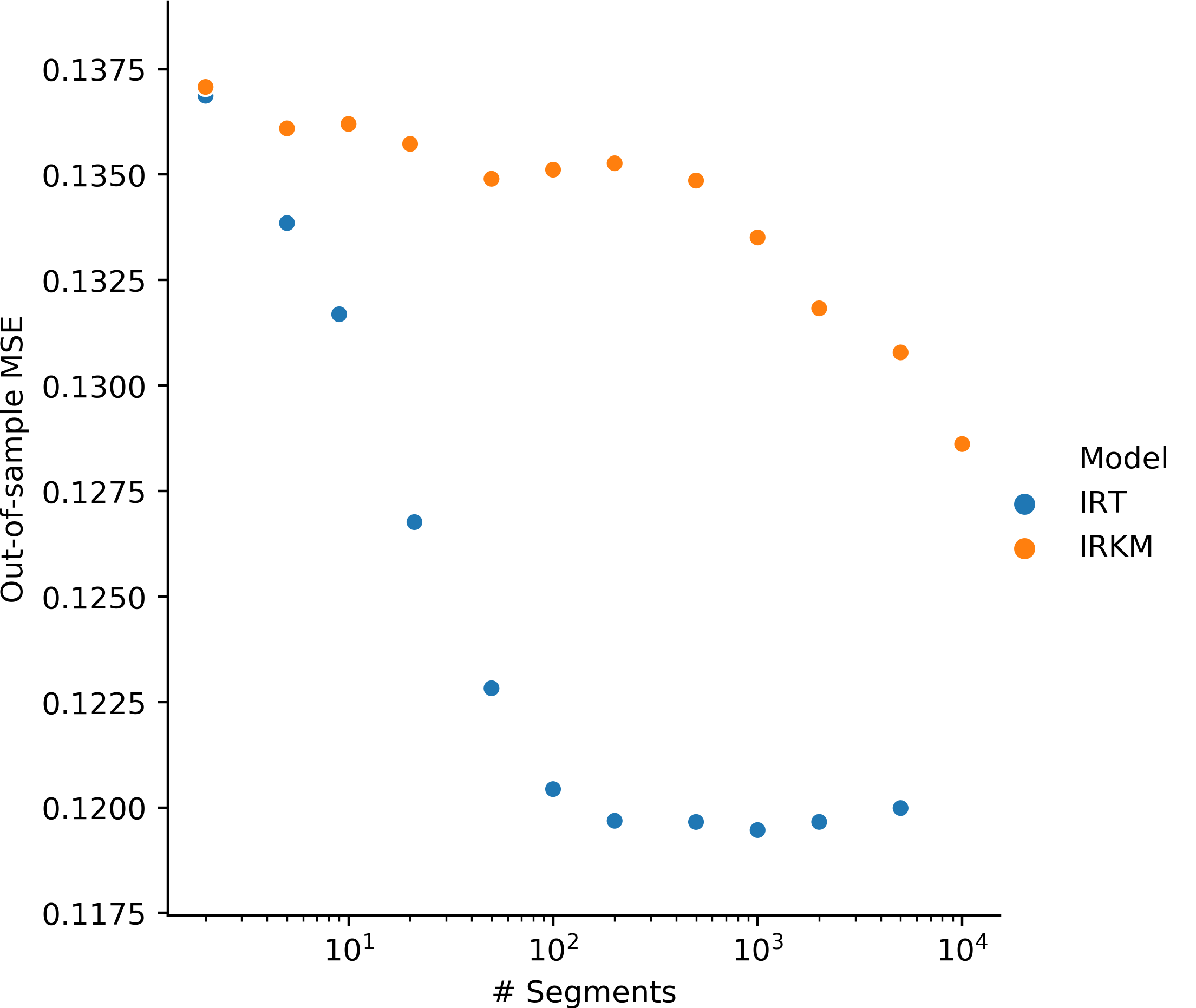}
\label{fig:mseexch3test}
\end{subfigure}
\begin{subfigure}{.49\linewidth}
\centering
\includegraphics[width=0.99\linewidth]{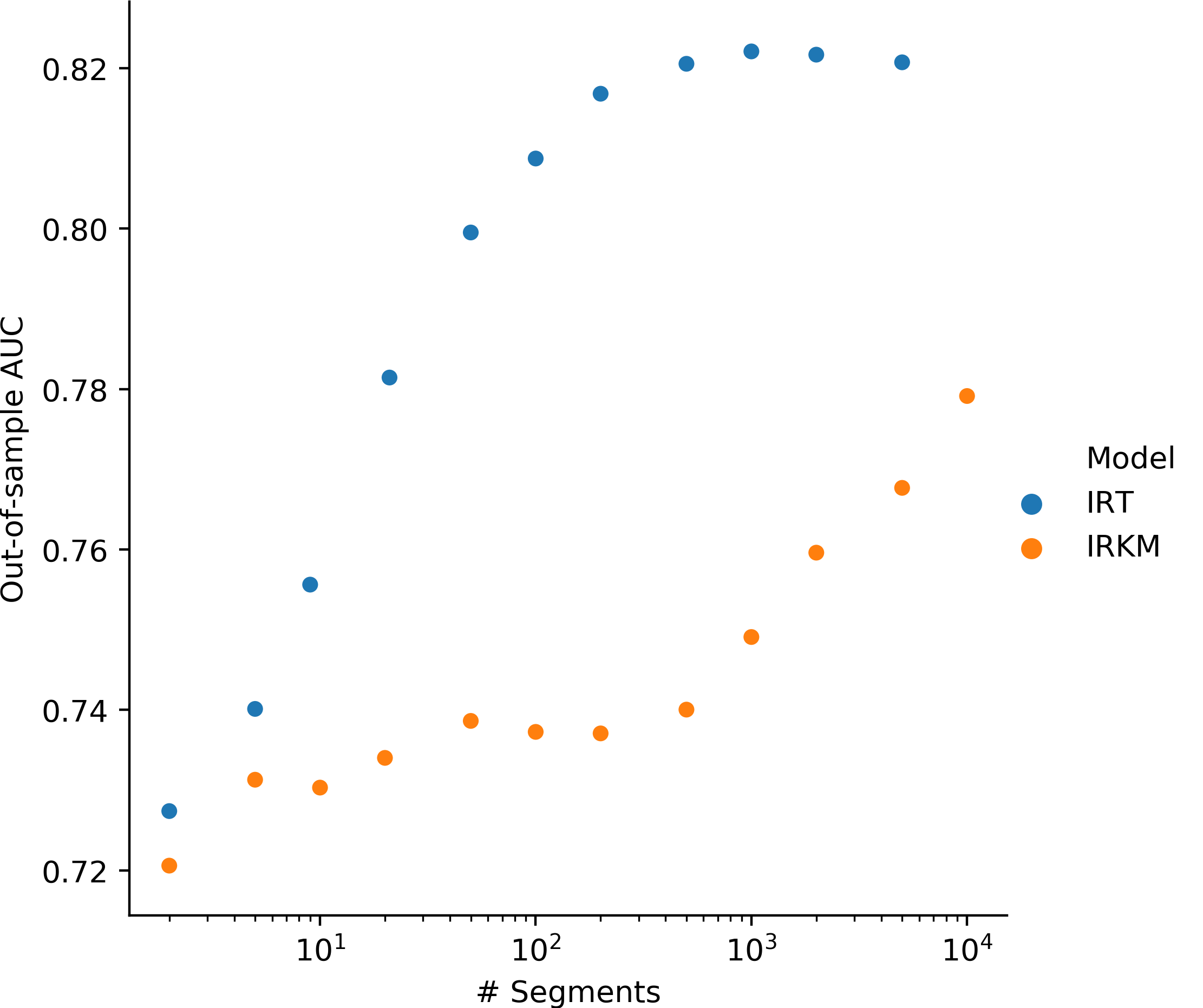}
\label{fig:aucexch3test}
\end{subfigure}
\par
	\medskip
	{\small  \textit{Note.} For IRT, we vary the number of segments from 1 to 5734. For IRKM, we vary the number of segments from 1 to 10000.}
\end{figure}

}


\clearpage \section{Example Splits Produced by the IRTs in Section \ref{sec:DSP}} \label{sec:DSPInt}

{\color{black}In Section \ref{sec:DSP} and Appendix \ref{sec:DSPExp}, we observe across all 3 ad exchanges that IRTs using only 10 market segments perform better than IRKM models using up to 1000 market segments. A key advantage of these concise IRTs is their interpretability, as their splits can be easily examined to determine which features have the greatest importance in customizing bids and predicting ad auction outcomes. To demonstrate this, Figure \ref{fig:irtsplits} visualizes the leaf models resulting from two different terminal splits found within the 10-segment IRTs trained on ad exchanges 1 and 2. Figure \ref{fig:channelsplit} corresponds to a terminal split on \textit{ad display channel} which assesses whether the display of the ad spot corresponds to desktop or mobile. The figure indicates that mobile display ads typically receive higher bids than desktop display ads, which could be due to mobile ads often having a higher click-through-rate (CTR) than desktop ads for this ad exchange. Figure \ref{fig:aspectratiosplit} visualizes the bid landscapes resulting from splitting on the value of the \textit{aspect ratio} of the ad spot (defined as the ratio of the width to the height of the ad spot). The figure suggests that aspect ratios greater than 3.5 typically receive lower bids than aspect ratios less than 3.5. This finding may be partially driven by aspect ratios greater than 3.5 also having predominately desktop displays - often with a lower CTR - as opposed to mobile displays (96\% have desktop displays for aspect ratios greater than 3.5 vs. 68\% for smaller aspect ratios). 
While IRTs can be used to study predictive links between features and response model differences (thereby aiding personalization), it is important to note that IRTs alone cannot be used to determine \textit{causal} relationships between features and differences in response models. 

\begin{figure}[h]
\FIGURE
  {
	\centering
	\begin{subfigure}{.45\linewidth}
		\centering
		\includegraphics[width=0.92\linewidth]{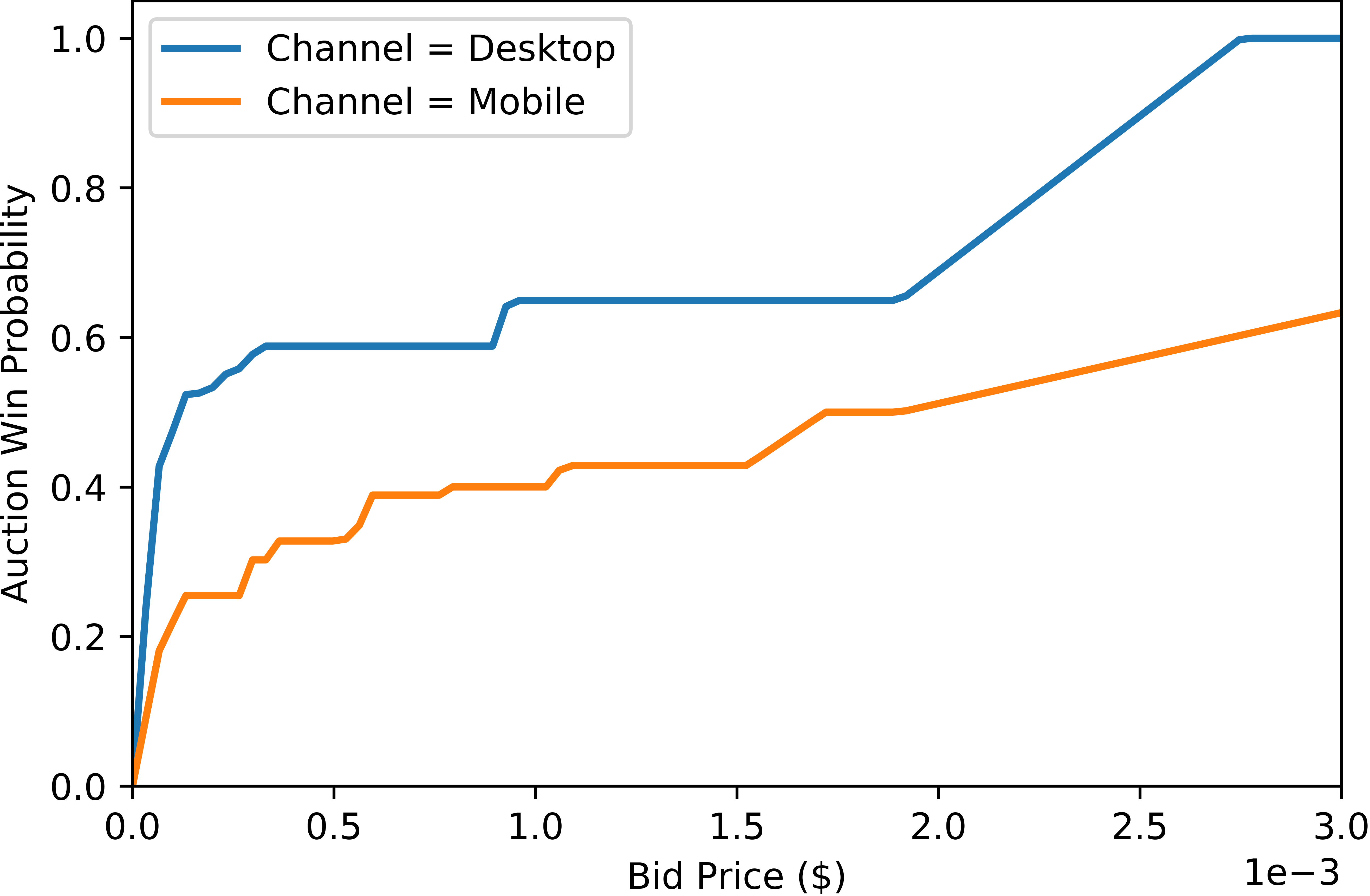}
		\caption{\small IR models from channel split}
		\vspace{0.1cm}
		\label{fig:channelsplit}
	\end{subfigure}%
	\begin{subfigure}{.45\linewidth}
		\centering
		\includegraphics[width=0.92\linewidth]{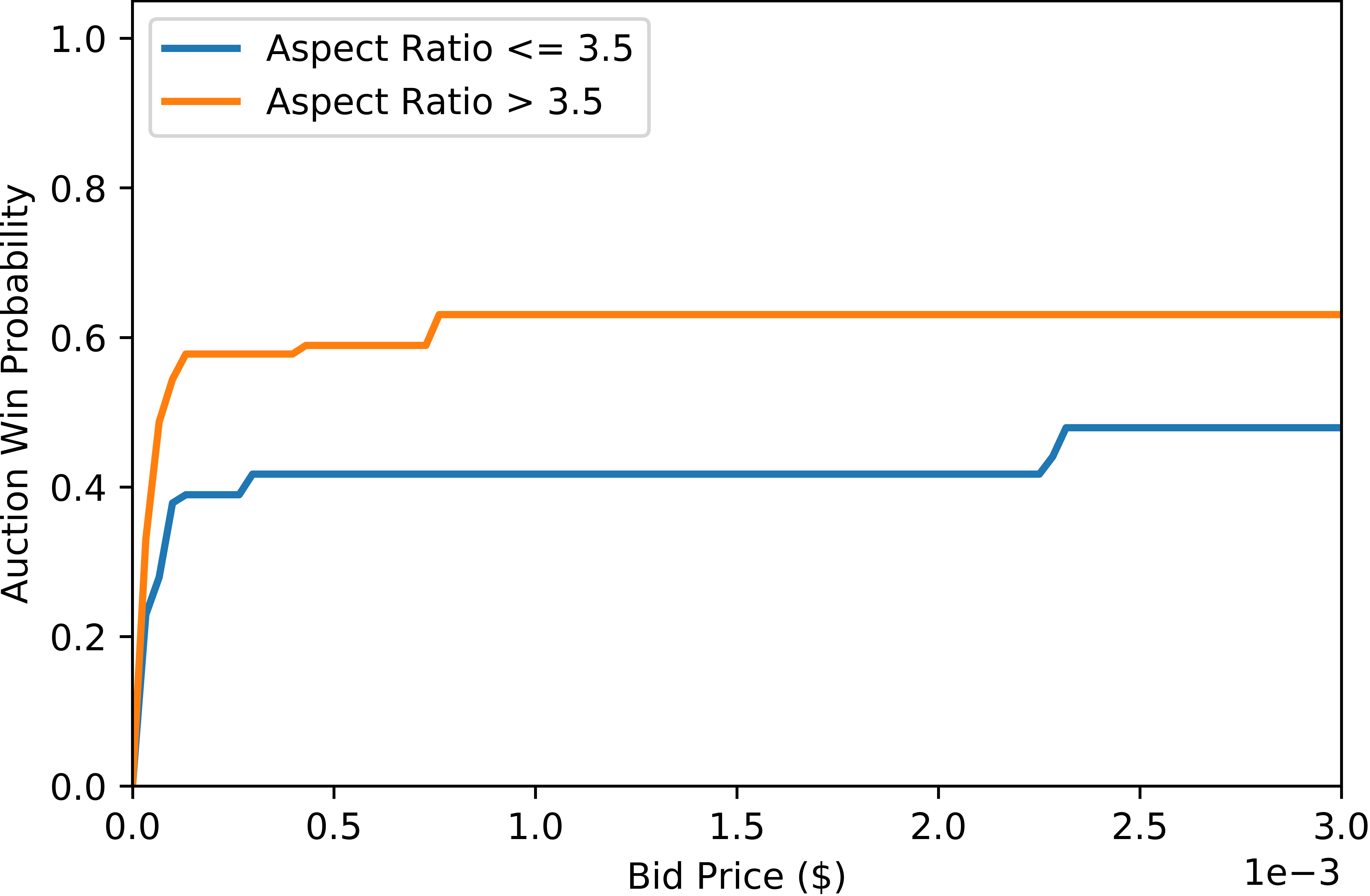}
		\caption{\small IR models from aspect ratio split}
		\vspace{0.1cm}
		\label{fig:aspectratiosplit}
	\end{subfigure}
	}
{Estimated bid landscapes from splits in the trained IRTs from Section \ref{sec:DSP}. \label{fig:irtsplits}}
{Figure \ref{fig:channelsplit} is from an IRT pruned to 10 leaves trained on bidding data from Exchange 2. Figure \ref{fig:aspectratiosplit} is from an IRT pruned to 10 leaves trained on bidding data from Exchange 1.}
\end{figure}
}


\end{APPENDICES}





\end{document}